\documentclass[a4paper,fleqn]{cas-sc}

\usepackage[authoryear,longnamesfirst]{natbib}
\usepackage{xcolor}
\usepackage{comment}
\usepackage{amsmath}
\usepackage{mdframed}
\usepackage{amsthm}
\usepackage{amsfonts}
\usepackage{pifont}
\usepackage{graphicx} 
\usepackage{tikz}
\usetikzlibrary{circuits.logic.IEC}
\usepackage{circuitikz}
\tikzstyle{union} = [rounded corners,text centered, draw=black]
\tikzstyle{intersection} = [rounded corners,text centered, draw=black]
\tikzstyle{times} = [circle,text centered, draw=black]
\tikzstyle{plus} = [circle,text centered, draw=black]
\tikzstyle{int} = [circle, text centered, draw=black]
\tikzstyle{leaf} = [rounded corners, text centered, draw=none, fill=none] 
\tikzstyle{arrow} = [thick,=,>=stealth]
\tikzstyle{dirarrow} = [thick,->,>=stealth]
\usetikzlibrary{arrows,arrows.meta,backgrounds,decorations.pathmorphing,positioning,fit,trees,shapes,shadows,automata,calc}
\usepackage{booktabs} 
\usepackage{subfig}
\usepackage{makecell}

\usepackage{enumitem}   

\usepackage[section]{placeins} 

\theoremstyle{definition}
\newtheorem{example}{Example}[section]

\theoremstyle{definition}
\newtheorem{definition}{Definition}[section]

\theoremstyle{definition}
\newtheorem{remark}{Remark}[section]

\theoremstyle{definition}
\newtheorem{proposition}{Proposition}[section]

\newcommand{\M}{\ensuremath{\mathcal{M}}}

\def\checkmark{\tikz\fill[scale=0.4](0,.35) -- (.25,0) -- (1,.7) -- (.25,.15) -- cycle;} 
\newcommand{\xmark}{\ding{55}}%

\newcommand{\WMC}{\text{WMC}} 

\newcommand{\new}[1]{\textcolor{DarkGreen}{#1}}

\def\tsc#1{\csdef{#1}{\textsc{\lowercase{#1}}\xspace}}
\tsc{WGM}
\tsc{QE}
\tsc{EP}
\tsc{PMS}
\tsc{BEC}
\tsc{DE}

\begin{document}
\let\WriteBookmarks\relax
\def\floatpagepagefraction{1}
\def\textpagefraction{.001}
\shorttitle{Weighted Model Counting and Probabilistic Model Checking}
\shortauthors{B. Salmani and V. Derkinderen.}

\title [mode = title]{Reasoning with Probabilities: Relating Weighted Model Counting and Probabilistic Model Checking}                      

\tnotetext[1]{This document is the results of the research
   project funded by KU Leuven.}

\author[1]{Bahare Salmani}[bioid=1,
                        orcid=0000-0003-3571-2502]
\cormark[1]
\ead{bahare.salmanibarzoki@kuleuven.de}

\credit{Conceptualization of this study, Methodology, Mappings, Writing}

\author[2]{Vincent Derkinderen}[bioid=2,
                        orcid=0000-0002-8894-270X]
\ead{vincent.derkinderen@kuleuven.be}

\affiliation[1-2]{organization={Department of Computer Science, KU Leuven},
                city={Leuven},
                country={Belgium}}

\cortext[cor1]{Corresponding author}


\begin{abstract}
Weighted model counting (WMC) and probabilistic model checking (PMC) are two well-established frameworks that are independently developed, the former for probabilistic inference, the latter traditionally for probabilistic verification, though recently also applied to inference. The formal relationship between the two frameworks, however, remains largely unexplored. In this paper, we lay the foundations for how they relate: we present (1) a mapping from cycle-free parametric Markov chains (pMCs) to arithmetic circuits (ACs), enabling the reduction of reachability probability computations in such pMCs to a weighted model counting problem on the corresponding ACs, and (2) a mapping from a subclass of arithmetic circuits — with probabilistic semantics — back to parametric Markov chains. We propose a detailed correspondence between the entities of WMC and PMC, and discuss how our mappings enable transferring optimization techniques such as bisimulation minimization across the frameworks.

\end{abstract}


\maketitle

\section{Introduction}
\label{Introduction}

\par \noindent \paragraph{Weighted Model Counting. } Weighted model counting \citep{DBLP:conf/aaai/SangBK05,DBLP:journals/ai/ChaviraD08,DBLP:conf/uai/DilkasB21} --and its algebraic extension \citep{DBLP:journals/japll/KimmigBR17,DBLP:conf/aaai/MaeneR25}-- is a dominant computational framework for probabilistic reasoning. It is used to determine the total weight of all satisfying assignments (models) of a logical formula, typically expressed in propositional logic. Each variable or literal in the formula is assigned a weight, and the weight of a model is calculated as the product of the weights of its literals. WMC extends traditional model counting, which simply counts the number of satisfying assignments, by incorporating probabilities or costs associated with each assignment. This makes it particularly useful in probabilistic reasoning, machine learning, and artificial intelligence to infer the likelihood of certain outcomes based on logical observations. Arithmetic circuits are the key underlying computation models in WMC.

\par \noindent \paragraph{Probabilistic Model Checking.} 
Probabilistic model checking \citep{DBLP:conf/lics/Katoen16} deals with automated analysis of stochastic systems. As in classical model checking, a state-based modelling approach is taken, which is extended with probabilities and non-determinism.
Typical models are Markov chains (MCs) and Markov Decision Processes (MDPs). More recently, partially-observable models have been also received attention~ 
\citep{DBLP:conf/atva/BorkJKQ20}. Similar to classical model checking, a key ingredient is a formal language for expressing \emph{properties}: probabilistic temporal logic such as Probabilistic Computation Tree Logic (PCTL) \citep{DBLP:journals/fac/HanssonJ94} is taken as a flexible and expressive language to express properties of interest.
A key procedure in verifying a MC against a temporal logic formula is to compute reachability probabilities, which amounts to computing the unique solution of a linear equation system whose size is linear in the number of states. Parametric Markov models are extensions of Markov models that allow polynomials over unknown parameters as transition probabilities. Over the last few decades, extensive synthesis techniques \citep{DBLP:conf/cav/DehnertJJCVBKA15,DBLP:conf/atva/QuatmannD0JK16,DBLP:conf/atva/GainerHS18,DBLP:conf/icse/FangCGA21,DBLP:conf/vmcai/HeckSJMK22,DBLP:journals/corr/abs-1903-07993,DBLP:conf/sefm/ElderhalliVHKT19} have been developed for such parametric models.
\par \noindent Probabilistic model checking and synthesis techniques built on it are applied in diverse domains such as dependability engineering and reliability analysis, see e.g., \citep{DBLP:journals/tii/VolkJK18,DBLP:conf/apn/JungesKS018,DBLP:journals/ress/GhadhabJKKV19}. The applicability of such techniques has recently been shown for inference, sensitivity analysis, and parameter tuning in Bayesian networks~\citep{DBLP:conf/qest/SalmaniK20,DBLP:conf/ecsqaru/SalmaniK21,DBLP:journals/jair/SalmaniK23}. While these studies provide extensive experimental results demonstrating the effectiveness of their methods in both parametric and non-parametric settings, they lack a comparative conceptual view on how probabilistic model checking differs from, or relates to, weighted-model-counting-based methods in modelling and computing probability distributions and (conditional) probabilities possibly beyond Bayesian networks.

   \begin{figure}
  \centering
  \begin{minipage}{0.01\linewidth}
  \end{minipage}
    \hfill
     \begin{minipage}{0.45\textwidth}
	 \centering
   \resizebox{0.75\width }{0.75\height}{%
  \subfloat[]{   \begin{tikzpicture} [node distance=1.1cm]
%

\node (n0) [plus] {$+$};

\node (n10) [times, below of=n0, left of=n0, xshift=-1.2cm] {$\times$};
\node (n11) [times, below of=n0, right of=n0, xshift=1.2cm] {$\times$};

\node (n20) [plus, below of=n10] {$+$};
\node (n21) [leaf, left of=n20] {$x_c$};
\node (n22) [leaf, right of=n20] {$c$};

\node (n25) [plus, below of=n11] {$+$};
\node (n24) [leaf, right of=n25] {$1-{x_c}$};
\node (n23) [leaf, left of=n25] {$\bar{c}$};

\node (n30) [times, below of=n20, left of=n20, xshift=0.25cm] {$\times$};
\node (n31) [times, below of=n20, right of=n20, xshift=-0.25cm] {$\times$};

\node (n32) [times, below of=n25, left of=n25, xshift=0.25cm] {$\times$};
\node (n33) [times, below of=n25, right of=n25, xshift=-0.25cm] {$\times$};

\node (n41) [leaf, below of=n30] {$s$};
\node (n40) [leaf, left of=n41, xshift=0.25cm] {$x_{s|c}$};
\node (n4r1) [plus, below of=n20, yshift=-1.1cm] {$+$};

\node (n43) [leaf, below of=n31] {$\bar{s}$};
\node (n42) [leaf, right of=n43, xshift=-0.25cm] {$1-x_{s|c}$};

\node (n45) [leaf, below of=n32] {$s$};
\node (n44) [leaf, left of=n45, xshift=0.25cm] {$x_{s|\bar{c}}$};
\node (n4r2) [plus, below of=n25, yshift=-1.1cm] {$+$};

\node (n47) [leaf, below of=n33] {$\bar{s}$};
\node (n46) [leaf, right of=n47] {$1-x_{s|\bar c}$};

\node (n50) [times, below of=n4r1, left of=n4r1, xshift=0.25cm] {$\times$};
\node (n51) [times, below of=n4r1, right of=n4r1, xshift=-0.25cm] {$\times$};
\node (n52) [times, below of=n4r2, left of=n4r2, xshift=0.25cm] {$\times$};
\node (n53) [times, below of=n4r2, right of=n4r2, xshift=-0.25cm] {$\times$};

\node (n60) [leaf, below of=n50, left of=n50, xshift=0.7cm] {$x_{r|c}$};
\node (n61) [leaf, below of=n50, right of=n50, xshift=-0.7cm] {$r$};

\node (n62) [leaf, below of=n51, left of=n51, xshift=0.7cm] {$ 1-x_{r|c}$};
\node (n63) [leaf, below of=n51, right of=n51, xshift=-0.6cm] {$\bar{r}$};

\node (n64) [leaf, below of=n52, left of=n52, xshift=0.7cm] {$x_{r|\bar{c}}$};
\node (n65) [leaf, below of=n52, right of=n52, xshift=-0.7cm] {$r$};

\node (n66) [leaf, below of=n53, left of=n53, xshift=0.7cm] {$1-{x_{r|\bar{c}}}$};
\node (n67) [leaf, below of=n53, right of=n53, xshift=-0.6cm] {$\bar{r}$};

    \draw [dirarrow] (n10) -- (n0);
    \draw [dirarrow] (n11) -- (n0);

    \draw [dirarrow] (n20) -- (n10);
    \draw [dirarrow] (n21) -- (n10);
    \draw [dirarrow] (n22) -- (n10);

    \draw [dirarrow] (n23) -- (n11);
    \draw [dirarrow] (n24) -- (n11);
    \draw [dirarrow] (n25) -- (n11);

    \draw [dirarrow] (n30) -- (n20);
    \draw [dirarrow] (n31) -- (n20);
    \draw [dirarrow] (n32) -- (n25);
    \draw [dirarrow] (n33) -- (n25);

    \draw [dirarrow] (n40) -- (n30);
    \draw [dirarrow] (n41) -- (n30);
    \draw [dirarrow] (n42) -- (n31);
    \draw [dirarrow] (n43) -- (n31);
    \draw [dirarrow] (n4r1) -- (n30);
    \draw [dirarrow] (n4r1) -- (n31);

    \draw [dirarrow] (n44) -- (n32);
    \draw [dirarrow] (n45) -- (n32);
    \draw [dirarrow] (n46) -- (n33);
    \draw [dirarrow] (n47) -- (n33);
    \draw [dirarrow] (n4r2) -- (n32);
    \draw [dirarrow] (n4r2) -- (n33);

    \draw [dirarrow] (n50) -- (n4r1);
    \draw [dirarrow] (n51) -- (n4r1);
    \draw [dirarrow] (n52) -- (n4r2);
    \draw [dirarrow] (n53) -- (n4r2);

    \draw [dirarrow] (n60) -- (n50);
    \draw [dirarrow] (n61) -- (n50);
    \draw [dirarrow] (n62) -- (n51);
    \draw [dirarrow] (n63) -- (n51);

    \draw [dirarrow] (n64) -- (n52);
    \draw [dirarrow] (n65) -- (n52);
    \draw [dirarrow] (n66) -- (n53);
    \draw [dirarrow] (n67) -- (n53);

\end{tikzpicture}
  }
     }
     \end{minipage}
    \centering
    \hfill
	 \begin{minipage}{0.45\textwidth}
	 \centering
         \resizebox{0.75\width }{0.75\height}{%
         \subfloat[]{
        \begin{tikzpicture}[
		node distance=0.3cm and 0.3cm,
		mNode/.style={draw,ellipse,align=center, minimum size=0.5cm},
		mLNode/.style={align=center, minimum size=0.5cm}
		]
		\node[](dummy){};
		\node[mNode,label=left:{}][below=0.4cm of dummy] (init) {$s_0$}; 
		\node[mNode,label=above left:{$c$}][below left=1.cm and 0.8cm of init] (s1) {$s_1$};
		\node[mNode,label=above right:{$\bar{c}$}][below right=1.cm and 0.8cm of init] (s2){$s_2$}; 
		
		\node[mNode,below=0.8cm of s1,label=above left:{$\bar{s}$}] (s4){$s_4$};
		\node[mNode,left=1.6cm of s4,label=above:{$s$}] (s3) {$s_3$};
		\node[mNode,below=0.8cm of s2,label=above left:{$ s$}] (s5){$s_5$};
		\node[mNode,right=1.6cm of s5, label=above:{$\,\,\,
        \,\,\, \bar{s}$}] (s6){$s_6$};
        \node[mNode,label=left:{$r$}][below=1.5cm of s4] (s11) {$s_{11}$};
              \node[mNode,label=right:{$\bar{r}$}][below=1.5cm of s5] (s12){$s_{12}$};

		\draw [-{Latex[length=2mm]}] (dummy) -- (init) node[midway, above, sloped] {};
		\draw [-{Latex[length=2mm]}] (init) -- (s1) node[midway, above, sloped] {$x_c$};
		\draw [-{Latex[length=2mm]}] (init) -- (s2) node[midway, above, sloped] {$1-x_c$};
		
		\draw [-{Latex[length=2mm]}] (s1) -- (s4) node[midway,anchor=south west, right] {$1-x_{s|c}$};
		\draw [-{Latex[length=2mm]}] (s1) -- (s3) node[midway, above, sloped] {$x_{s|c}$};
		\draw [-{Latex[length=2mm]}] (s2) -- (s5) node[midway, anchor=south west, right] {$x_{s|\bar{c}}$};
		\draw [-{Latex[length=2mm]}] (s2) -- (s6) node[pos=0.4, above, sloped] {$1{-}x_{s|\bar{c}}$};	
		
		\draw [-{Latex[length=2mm]}] (s4) -- (s12) node[pos=0.25,yshift=-2.5pt, above, sloped] {$1-x_{r|c}$}; //s4 ->
		\draw [-{Latex[length=2mm]}] (s4) -- (s11) node[pos=0.25, yshift=-2pt,above, sloped] {$x_{r|c}$}; //s4 ->
		\draw [-{Latex[length=2mm]}] (s3) -- (s11) node[pos=0.25, left] {$x_{r|c}$}; //s3 ->
		\draw [-{Latex[length=2mm]}] (s3) -- (s12) node[pos=0.3,yshift=-2pt, above, sloped] {$1{-}x_{r|c}$};	 //s3->
		\draw [-{Latex[length=2mm]}] (s5) -- (s12) node[pos=0.5, yshift=-2pt,above, sloped] {$1{-}x_{r|\bar{c}}$}; // s5->
		\draw [-{Latex[length=2mm]}] (s5) -- (s11) node[pos=0.2,yshift=-2pt, above, sloped] {$x_{r|\bar{c}}$}; //s5 ->
		\draw [-{Latex[length=2mm]}] (s6) -- (s11) node[pos=0.15, above, sloped] {$x_{r|\bar{c}}$};
		\draw [-{Latex[length=2mm]}] (s6) -- (s12) node[pos=0.4,sloped, below] {$1{-}x_{r|\bar{c}}$};
		
		\path[every loop/.append style=-{Latex[length=2mm]}]
		(s11) edge [loop below] node {1} (s11)
		(s12) edge [loop below] node {1} (s12);
	\end{tikzpicture}
	}
    }
     \end{minipage}
     \centering
     \hfill
   \begin{minipage}{0.02\linewidth}
  \end{minipage}
     \caption{An Arithmetic circuit and its corresponding cycle-free parameteric Markov chain.}
\label{fig-mc-ac-updated}
\end{figure}
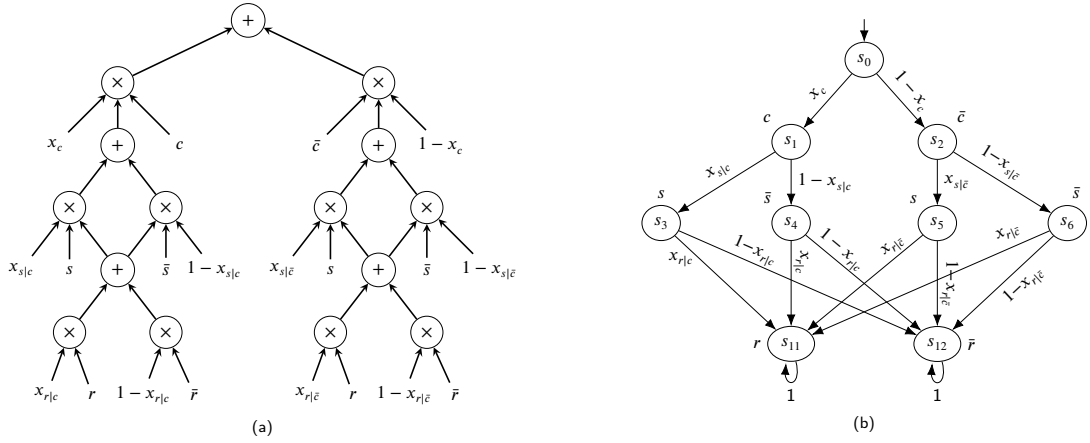

\par \noindent \paragraph{Main Objective.} Our aim is to explore conceptual bridges between probabilistic model checking and weighted model counting, particularly in the context of probabilistic inference.
\par \noindent \paragraph{Contributions. } Our main contributions can be listed as follows.
\begin{itemize}
  \item We present a formal mapping from cycle-free parametric MCs to arithmetic circuits that enables reducing the computation of reachability probability in such pMCs to a weighted model counting on the corresponding circuits. Fig. \ref{fig-mc-ac-updated}, for example, shows a pMC (right) alongside its AC (left) obtained from our mapping. 
  \item We present a formal mapping from certain subclasses of ACs---with probabilistic semantics---to pMCs, thereby reducing circuit evaluation to computing reachability probabilities in pMCs. 
  \item We discuss how the notions from probabilistic model checking connect to those from weighted model counting. 
    \item We illustrate, through examples on Bayesian networks, how probabilistic bisimulation---an equivalence relation used to simplify Markov models---relates to (conditional) independence in the underlying probability distribution. This offers a semantic explanation for why bisimulation minimization yields significant inference-time improvements when probabilistic model checking is applied to Bayesian networks in previous literature. 
  \end{itemize}
\par \noindent 

\par \noindent Note that since our mappings preserve the structure of the input formalism, they suggest that tractability results established for one framework can transfer to the other: provided that the conditions known to enable efficient weighted model counting (e.g., determinism or decomposability of the underlying circuit) can be ensured from the structural properties of the corresponding Markov chain.
\vspace{0.2cm}
\par \noindent We emphasize that the existing literature does not provide a direct comparative study between weighted model counting and probabilistic model checking in general, focusing only on Bayesian networks. The formal mappings and semantic links presented here go beyond this: our results apply not only to Bayesian networks but to arbitrary acyclic probabilistic (logic) programs. While some of the ideas and examples in this document were first introduced in the dissertation \citep{salmani2024probabilistic}, these formal mappings and semantic links are original to this work.


\section{Preliminaries}
\label{sec:prel}

We cover in this section, the preliminaries related to weighted model counting and probabilistic model checking.
\subsection{Propositional Logic}
In propositional logic, symbols are used to represent atomic propositions; e.g., symbol $r$ can denote that it is raining. Such a proposition can be negated: $\bar r$ denotes the negation. A \emph{literal} is an atomic proposition or its negation. These atomic propositions can be combined to form more complex formulas. A \emph{logic formula} $\phi$ in propositional logic is inductively defined as an atom $\phi$, the negation of a formula $\neg \phi_1$, the conjunction of two formulas $\phi_1 \land \phi_2$ (read as `and'), or the disjunction of two formulas $\phi_1 \lor \phi_2$ (read as `or'). A \emph{logic circuit} also represents a logic formula, with the distinction that it explicitly models the reuse of subformulas. E.g., $(\phi_1 \land r) \lor (\phi_1 \land c)$ is logic circuit where $\phi_1$ is used twice. We use formula and logic circuit interchangeably.

A truth assignment to each atomic proposition---also called an \emph{interpretation}---induces a truth value for each logic circuit over those propositions. For example, if we assume that it is raining ($r$) and that it is cloudy ($c$), then we also induce a truth value for $r \land c$. An interpretation $I$ may be represented as a set of literals, such that $a \in I$ if atomic proposition $a$ is considered true, and $\bar a \in I$ if it is considered false.
A \emph{model} of a logic circuit $\phi$ is an interpretation for which $\phi$ is induced to be true. We will use $\mathcal{R}(\phi)$ to refer to the set of models of $\phi$, thereby implicitly assuming a set of atomic propositions such that $\mathcal{R}(\phi)$ is finite.

\begin{example}
    Consider atomic propositions $r$ and $c$. The interpretation $I = \{r, \bar c\}$ is one of the models for the logic circuit $\phi = r \lor c$. In total, there are three models; i.e., $|\mathcal{R}(\phi)| = 3$.
\end{example}

A variety of computational tasks exist over a logic circuit $\phi$, some of which we touch upon later in this work. Important, is that the structure of $\phi$ influences the tractability of a computational task. In this work, we are particularly interested in logic circuits that are in \emph{deterministic decomposable negation normal form} (d-DNNF)~\citep{Darwiche02KCMap}.

\begin{definition}[determinism]
    A logic circuit $\phi$ is deterministic (d) iff for each $\phi_1 \lor \phi_2$, the subcircuits $\phi_1$ and $\phi_2$ do not share any models. I.e., $\mathcal{R}(\phi_1) \cap \mathcal{R}(\phi_2) = \emptyset$.
\end{definition}

\begin{definition}[decomposability]
    A logic circuit $\phi$ is decomposable (D) iff for each $\phi_1 \land \phi_2$, the subcircuits $\phi_1$ and $\phi_2$ do not share any atomic propositions.
\end{definition}

\begin{definition}[negation normal form]
    A logic circuit $\phi$ is in negation normal form (NNF) iff negation only occurs over the atomic propositions in $\phi$. In other words, iff the logic circuit is a combination of literals using the $\land$ and $\lor$ connectives.
\end{definition}


\subsection{Random Variables and Parametric Probability Distributions}
Let $X \,{=}\, \{x_1, \ldots, x_n\}$ be a set of $n$ real-valued parameters.
A \emph{parameter instantiation} is a function $u\,{:}\, X \,{\rightarrow}\,\mathbb{R}$ that maps each parameter to a value.
Parametric probability distributions are extensions of probability distributions, where probability values are not constant probabilities, but rather polynomials with rational coefficients. 
Let $\mathbb{Q}[X]$ be the set of multivariate polynomials with rational coefficients over $X$.
Formally, a \emph{parametric probability distribution} \new{on a finite set $\Omega$} is the function $\mu: \Omega \rightarrow \mathbb{Q}[X]$ with $\sum_{\omega \in \Omega} \mu(\omega) \,{=}\, 1$. Let $\mbox{\it pDistr}(\Omega)$ denote the set of {parametric} probability distributions over $\Omega$ with parameters in $X$. Instantiation $u$ is \emph{well-formed} for $\mu \,{\in}\, \mbox{\it pDistr}(\Omega)$ iff $0 \leq \mu(\omega)[u] \leq 1$ and $\Sigma_{\omega \in \Omega} \, \mu(\omega)[u] = 1$. In this paper, we assume that all the instantiations are well-formed in the parametric setting.

\section{Models}
In this section, we overview the models relevant to this paper: arithmetic circuits and (parametric) Markov chains.  
\subsection{Arithmetic Circuits}
\label{sec:wmc}
Arithmetic circuits in general provide a representation for polynomials~\citep{DBLP:conf/kr/Darwiche02,DBLP:journals/jacm/Darwiche03}. In the context of probabilistic inference, they provide a tractable \emph{computational representation} for probability distributions.

\begin{definition}[Arithmetic Circuit]
    An arithmetic circuit (AC) is a single-rooted directed acyclic graph $\mathcal{C} = (N, n_0, \mbox{\it children},op,sym)$ over the parameters $X$ and the literal set $Lits$; it is comprised of a set of nodes $N$ that we partition into leaf nodes $N_L$ and inner nodes $N_I$, a root node $n_0 \in N$, a function $\mbox{\it children} : N \rightarrow 2^N$ that denotes the set of children for each node, a function $op: N_I \rightarrow \{+, \times\}$ that associates each inner node with an arithmetic operation, and a function $sym : N_L \rightarrow (\mathbb{Q}[X] \cup Lits)$ that associates each leaf node with either a constant $c \in \mathbb{Q}$, or a polynomial $f \in \mathbb{Q}[X]$, or a literal $l \in Lits$.
\end{definition}

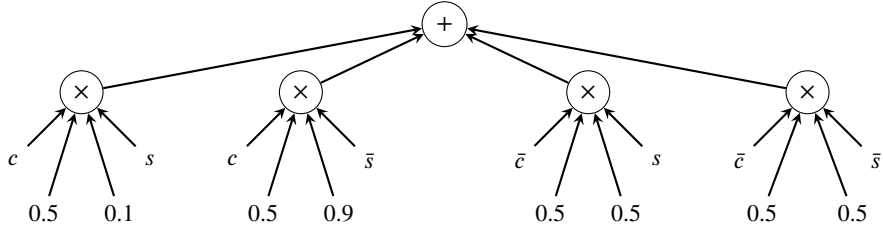
\begin{figure}
    \centering
    \begin{tikzpicture} [node distance=0.9cm]
%

\node (n0) [plus] {$+$};

\node (n10) [times, below of=n0, left of=n0, xshift=-1cm] {$\times$};
\node (n11) [times, below of=n0, right of=n0, xshift=1cm] {$\times$};

\node (n12) [times, left of=n10, xshift=-2cm] {$\times$};
\node (n13) [times, right of=n11, xshift=2cm] {$\times$};

\node (n200) [leaf, below of=n12, left of=n12] {$c$};
\node (n201) [leaf, below of=n12, right of=n12] {$s$};
\node (n202) [leaf, below of=n12, left of=n12, xshift=0.4cm, yshift=-0.7cm] {$0.5$};
\node (n203) [leaf, below of=n12, right of=n12, xshift=-0.4cm, yshift=-0.7cm] {$0.1$};

\node (n210) [leaf, below of=n10, left of=n10] {$c$};
\node (n211) [leaf, below of=n10, right of=n10] {$\bar{s}$};
\node (n212) [leaf, below of=n10, left of=n10, xshift=0.4cm, yshift=-0.7cm] {$0.5$};
\node (n213) [leaf, below of=n10, right of=n10, xshift=-0.4cm, yshift=-0.7cm] {$0.9$};

\node (n220) [leaf, below of=n11, left of=n11] {$\bar{c}$};
\node (n221) [leaf, below of=n11, right of=n11] {$s$};
\node (n222) [leaf, below of=n11, left of=n11, xshift=0.4cm, yshift=-0.7cm] {$0.5$};
\node (n223) [leaf, below of=n11, right of=n11, xshift=-0.4cm, yshift=-0.7cm] {$0.5$};

\node (n230) [leaf, below of=n13, left of=n13] {$\bar{c}$};
\node (n231) [leaf, below of=n13, right of=n13] {$\bar{s}$};
\node (n232) [leaf, below of=n13, left of=n13, xshift=0.3cm, yshift=-0.7cm] {$0.5$};
\node (n233) [leaf, below of=n13, right of=n13, xshift=-0.3cm, yshift=-0.7cm] {$0.5$};

    \draw [dirarrow] (n10) -- (n0);
    \draw [dirarrow] (n11) -- (n0);
    \draw [dirarrow] (n12) -- (n0);
    \draw [dirarrow] (n13) -- (n0);

    \draw [dirarrow] (n200) -- (n12);
    \draw [dirarrow] (n201) -- (n12);
    \draw [dirarrow] (n202) -- (n12);
    \draw [dirarrow] (n203) -- (n12);

    \draw [dirarrow] (n210) -- (n10);
    \draw [dirarrow] (n211) -- (n10);
    \draw [dirarrow] (n212) -- (n10);
    \draw [dirarrow] (n213) -- (n10);

    \draw [dirarrow] (n220) -- (n11);
    \draw [dirarrow] (n221) -- (n11);
    \draw [dirarrow] (n222) -- (n11);
    \draw [dirarrow] (n223) -- (n11);

    \draw [dirarrow] (n230) -- (n13);
    \draw [dirarrow] (n231) -- (n13);
    \draw [dirarrow] (n232) -- (n13);
    \draw [dirarrow] (n233) -- (n13);
\end{tikzpicture}
  
    \caption{An arithmetic circuit that represents a polynomial, and that models a probability distribution of the form $Pr(S,C) = Pr(S|C)Pr(C)$ with $Pr(c) = 0.5$, $Pr(s|c) = 0.1$, and $Pr(s|\neg c) = 0.5$.} 
    \label{fig:ac-example-small}
\end{figure}

\begin{example} Figure~\ref{fig:ac-example-small} illustrates an arithmetic circuit $\mathcal{C}$ over the literal set $Lits = \{c, \neg c, s, \neg s\}$ that corresponds to polynomial $f_{\mathcal{C}}$.
\begin{equation*}
    f_{\mathcal{C}} = 0.5 \cdot 0.1 \cdot c \cdot s + 0.5 \cdot 0.9 \cdot c \cdot \neg s + 0.5 \cdot 0.5 \cdot \neg c \cdot s + 0.5 \cdot 0.5 \cdot \neg c \cdot \neg s
\end{equation*} 
The circuit represents a probability distribution as described in the figure's caption. We can evaluate an AC by (i) associating with each literal in the circuit a value in $\{0,1\}$ that represents that literal's truth value, and (ii) propagating those values upwards in the circuit. E.g., consider the evidence \emph{no sprinkler} that is mapped to the literal $\neg s$. To compute $\Pr(\neg s)$, the circuit is evaluated by the weight function $w$ with $w(\neg s) = 1$, $w(s) = 0$, $w(\neg c) = 1$, and $w(c) = 1$: the literals appearing in the evidence are weighted by $1$, their negation is weighted by $0$, and the literals of the atomic propositions not appearing in the evidence are also all weighted by $1$. We obtain the circuit evaluation $\mathcal{C}[w] = 0.5 \cdot 0.1 \cdot 0 + 0.5 \cdot 0.9 \cdot 1 + 0.5 \cdot 0.2 \cdot 0 + 0.5 \cdot 0.5 \cdot 1 = 0.7$.

\end{example}

\begin{definition}[Circuit Evaluation] Given an AC $\mathcal{C} = (N,n_0,\mbox{\it children},op,sym)$ over the literals $Lits$, and an assignment function $w$ that maps each literal $l$ to its assigned value $w(l) \in \mathbb{Q}$ or the polynomial $f \in \mathbb{Q}[X]$, the circuit evaluation $\mathcal{C}(n)[w]$ is defined inductively for each node $n$ as follows. 
    \begin{equation}
    \label{eq-circuit-evaluation}
        \mathcal{C}(n)[w] =
        \begin{cases}
            sym(n), \quad &\text{ if $n$ is a leaf node and $sym(n)$ is a constant} \\
            w(sym(n)), \quad &\text{ if $n$ is a leaf node and $sym(n)$ is a literal} \\
            \sum_{n_i \in \mbox{\it children}(n)} \mathcal{C}(n_i)[w], \quad &\text{ if $n$ is an inner node and $op(n) = +$} \\
            \prod_{n_i \in \mbox{\it children}(n)} \mathcal{C}(n_i)[w]. \quad &\text{ if $n$ is an inner node and $op(n) = \times$}
        \end{cases}
    \end{equation}
\end{definition}

\par \noindent We are often interested in $\mathcal{C}(n_0)[w]$, i.e., the circuit evaluation at its root node and simply denote it by $\mathcal{C}[w]$. Not every arithmetic circuit represents a probability distribution. \cite{DBLP:journals/ai/ChaviraD08} have demonstrated how to construct a correct arithmetic circuit from a Bayesian network, enabling exact inference tractable in the size of the new representation. 
This construction is best explained through weighted model counting.

\begin{definition}[Weighted Model Count]
    Given a propositional logic circuit $\phi$ over a set of literals $Lits$, and a weight function $w: Lits \rightarrow \mathbb{R}$ that maps each literal to a weight, the weighted model count $\WMC(\phi,w)$ is defined as
    \begin{equation}\label{eq:wmc}
            \WMC(\phi,w) = \sum_{I \in \mathcal{R}(\phi)} \prod_{lit \in I} w(lit).
    \end{equation}
    This represents a weighted sum over all models $I \in \mathcal{R}(\phi)$, where we assume that the weight of a model $I$ factorizes into a factor per literal of that model: $\prod_{lit \in I} w(lit)$. Despite this assumption, we can use weighted model counting to perform inference on more general distributions, like those represented as a Bayesian network~\citep{DBLP:journals/jacm/Darwiche03}.
\end{definition}

\begin{example}[Weighted Model Count]
    Consider as example logic circuit $\phi = c \lor \neg s$, with weight function $w$ defined as $w(c) = w(\neg c) = 0.5$, $w(s) = 0.3$, and $w(\neg s) = 0.7$. There are three models satisfying $\phi$ with the weights $0.15$, $0.35$, and $0.35$. We thus have $\WMC(\phi,w) = 0.15 + 0.35 + 0.35 = 0.85$.
\end{example}

\paragraph{WMC to Arithmetic Circuits.}
Probabilistic inference can be cast as a weighted model counting task. By definition of the weighted model count (cf. Equation~\ref{eq:wmc}), this task forms a polynomial, meaning an arithmetic circuit can be used to conduct probabilistic inference.
As is, the equation is very inefficient however, as it aggregates each model one by one, and in the worst case there are $2^{|\mathbf{V}|}$ models. Fortunately we can exploit properties, like commutativity, associativity and distributivity, to rewrite this polynomial into a much more efficient form. 

Since the polynomial's structure is dictated by logic circuit $\phi$, a common approach is to use knowledge compilation techniques 
to transform $\phi$ into a logically equivalent circuit $\phi'$, from which a correct arithmetic circuit---one that correctly represents the weighted model count polynomial for $\phi$---can more easily be extracted.
In practice, this means transforming $\phi$ into a circuit $\phi'$ that is in deterministic decomposable negation normal form (d-DNNF). If $\phi'$ is a d-DNNF circuit, then we can construct a suitable arithmetic circuit using the following procedure: replace each $\lor$ and $\land$ with $+$ and $\times$ respectively, and then replace the propositional variable assignments in the leaf nodes with the appropriate weights of the weight function $w$. The resulting structure evaluates exactly to the weighted model count. 

\subsection{Parameteric Markov Chains}
\label{sec:pmc}

Markov chains and their extensions (with non-determinism) are key models in \emph{probabilistic model checking}. In this paper, we only consider reward-free deterministic Markov chains that are relevant for probabilistic inference. We base our definition on parametric Markov chains to distinguish between topology and concrete probability values \footnote{In this work, we do not emphasize on (1) the interesting properties of parametric Markov chains such as arbitrary parameter dependencies and (2) the synthesis problems and algorithms for finding parameter values based on constraints. }.

\begin{definition}[parametric Markov Chain]
A parametric Markov chain (pMC) ${\M}$ is a tuple $(S, s_0, X, \mbox{Lits}, L, P)$, 
where ${S}$ is a countable non-empty set of states, 
$s_0 \in S$ is the initial state, $X$ is a set of parameters,
${\mbox{Lits}}$ is a countable set of \emph{literals} \footnote{We use literals rather than the standard \emph{atomic propositions} in probabilistic model checking literature to make a common ground with weighted model counting literature. }, 
${L } \colon S \to 2^{\mbox{Lits}}$ is a {labeling function} that maps each state to a subset of literals,
and ${P} \colon S \to \mbox{pDistr}_X(S)$ is a \emph{transition probability function} that for each state $s$ in $S$ defines a 
\emph{parametric probability distribution} over the states.
\end{definition}
\par \noindent The transition probability of going from state $s$ to $s'$ is given by $P(s)(s')$ denoted as $P(s, s')$ that is a polynomial over $X$. 
A pMC with $X = \emptyset$ and $P \colon S \to Distr(S)$ is a Markov chain (MC). Applying a well-formed parameter instantiation $u \colon X \to \mathbb{R}$ on the pMC ${\M} = (S, s_0, X, \mbox{Lits}, L, P)$ induces the MC $\mathcal{M}[u]$ where $P[u](s)$, (i.e., the concrete value that is obtained by instantiating the parameters by their values from $u$) is a probability distribution over $S$. 
 \begin{figure}
     \centering
   \begin{minipage}{0.01\linewidth}
  \end{minipage}
  \hfill
    \centering
	 \begin{minipage}{0.47\textwidth}
	 \centering
    \resizebox{0.67\width }{0.67\height}{%
     \subfloat[]{
        \begin{tikzpicture}[
		node distance=0.3cm and 0.3cm,
		mNode/.style={draw,ellipse,align=center, minimum size=0.5cm},
		mLNode/.style={align=center, minimum size=0.5cm}
		]
		\node[](dummy){};
		\node[mNode,label=left:{}][below=0.5cm of dummy] (init) {$s_0$}; 
		\node[mNode,label=above left:{$c$}][below left=1.cm and 0.8cm of init] (t1) {$s_1$};
		\node[mNode,label=above right:{$\bar{c}$}][below right=1.cm and 0.8cm of init] (f1){$s_2$}; 
		
		\node[mNode,below=1cm of t1,label=left:{$c, \bar{s}$}] (t1t3){$s_4$};
		\node[mNode,left=1.6cm of t1t3,label=above:{$c, s$}] (t1f3) {$s_3$};
		\node[mNode,below=1cm of f1,label=left:{$\bar{c}, s$}] (f1t3){$s_5$};
		\node[mNode,right=1.6cm of f1t3, label=above:{$\,\,\,
        \,\,\,\bar{c}, \bar{s}$}] (f1f3){$s_6$};

		\node[mNode,below=1.6cm of t1f3, label=below:{$s, r$}] (t2) {$s_7$};
            \node[mNode,right=1.3cm of t2,label=right:{$s, \bar{r}$}] (t3){$s_8$};
		\node[mNode,below=1.6cm of f1f3, label=below:{$\,\,\,\,\,\,\,\,\,\,\,\,\,\,\,\,\,\,\bar{s}, \bar{r}$}] (f2) {$s_{10}$};
              \node[mNode,left=1.3cm of f2,label=right:{$\bar{s},r$}] (f3){$s_{9}$};
              
              	\node[mNode,label=left:{$w$}][below=2cm of t3] (s11) {$s_{11}$};
              \node[mNode,label=right:{$\bar{w}$}][below=2cm of f3] (s12){$s_{12}$};

		\draw [-{Latex[length=2mm]}] (dummy) -- (init) node[midway, above, sloped] {};
		\draw [-{Latex[length=2mm]}] (init) -- (t1) node[midway, above, sloped] {$x_c$};
		\draw [-{Latex[length=2mm]}] (init) -- (f1) node[midway, above, sloped] {$1-x_c$};
		
		\draw [-{Latex[length=2mm]}] (t1) -- (t1t3) node[midway,anchor=south west, right] {$1-x_{s|c}$};
		\draw [-{Latex[length=2mm]}] (t1) -- (t1f3) node[midway, above, sloped] {$x_{s|c}$};
		\draw [-{Latex[length=2mm]}] (f1) -- (f1t3) node[midway, anchor=south west, right] {$x_{s|\bar{c}}$};
		\draw [-{Latex[length=2mm]}] (f1) -- (f1f3) node[pos=0.4, above, sloped] {$1{-}x_{s|\bar{c}}$};	
		
		\draw [-{Latex[length=2mm]}] (t1t3) -- (f2) node[pos=0.6,yshift=-2.5pt, above, sloped] {$1-x_{r|c}$}; //s4 ->
		\draw [-{Latex[length=2mm]}] (t1t3) -- (f3) node[pos=0.75, yshift=-2pt,above, sloped] {$x_{r|c}$}; //s4 ->
		\draw [-{Latex[length=2mm]}] (t1f3) -- (t2) node[pos=0.25, left] {$x_{r|c}$}; //s3 ->
		\draw [-{Latex[length=2mm]}] (t1f3) -- (t3) node[pos=0.35,yshift=2.5pt, below, sloped] {$1{-}x_{r|c}$};	 //s3->
		\draw [-{Latex[length=2mm]}] (f1t3) -- (t3) node[pos=0.75, yshift=-2pt,above, sloped] {$1{-}x_{r|\bar{c}}$}; // s5->
		\draw [-{Latex[length=2mm]}] (f1t3) -- (t2) node[pos=0.55,yshift=-2pt, above, sloped] {$x_{r|\bar{c}}$}; //s5 ->
		\draw [-{Latex[length=2mm]}] (f1f3) -- (f3) node[pos=0.25, above, sloped] {$x_{r|\bar{c}}$};
		\draw [-{Latex[length=2mm]}] (f1f3) -- (f2) node[pos=0.4,sloped, above] {$1{-}x_{r|\bar{c}}$};

		\draw [-{Latex[length=2mm]}] (t2) -- (s11) node[pos=0.5, below, sloped] {$x_{w|s,r}$};
		\draw [-{Latex[length=2mm]}] (t2) -- (s12) node[pos=0.2, above,yshift=-2pt, sloped] {$1-x_{w|s,r}$};
		\draw [-{Latex[length=2mm]}] (t3) -- (s11) node[pos=0.4, yshift=-2pt,sloped,above] {$x_{w|s,\bar{r}}$};
		\draw [-{Latex[length=2mm]}] (t3) -- (s12) node[pos=0.35,yshift=-2pt, above, sloped] {$1-x_{w|s,\bar{r}}$};	
		\draw [-{Latex[length=2mm]}] (f2) -- (s11) node[pos=0.15, above,yshift=-2pt, sloped] {$x_{w|\bar{s}, \bar{r}}$};
		\draw [-{Latex[length=2mm]}] (f2) -- (s12) node[pos=0.45, below, sloped] {$1{-}x_{w|\bar{s}, \bar{r}}$};
        
		\draw [-{Latex[length=2mm]}] (f3) -- (s11) node[pos=0.15, above, sloped] {$x_{w|\bar{s}, r}$};
		\draw [-{Latex[length=2mm]}] (f3) -- (s12) node[pos=0.4, yshift=2pt,sloped,below] {$1{-}x_{w|\bar{s}, r}$};
		
		\path[every loop/.append style=-{Latex[length=2mm]}]
		(s11) edge [loop below] node {1} (s11)
		(s12) edge [loop below] node {1} (s12);
	\end{tikzpicture}
	}
    }
\end{minipage}
\hfill
 \begin{minipage}{0.47\textwidth}
	 \centering
    \resizebox{0.67\width }{0.67\height}{%
     \subfloat[]{
        \begin{tikzpicture}[
		node distance=0.3cm and 0.3cm,
		mNode/.style={draw,ellipse,align=center, minimum size=0.5cm},
		mLNode/.style={align=center, minimum size=0.5cm}
		]
		\node[](dummy){};
		\node[mNode,label=left:{}][below=0.5cm of dummy] (init) {$s_0$}; 
		\node[mNode,label=above left:{$c$}][below left=1.cm and 0.8cm of init] (t1) {$s_1$};
		\node[mNode,label=above right:{$\bar{c}$}][below right=1.cm and 0.8cm of init] (f1){$s_2$}; 
		
		\node[mNode,below=1cm of t1,label=left:{$c, \bar{s}$}] (t1t3){$s_4$};
		\node[mNode,left=1.6cm of t1t3,label=above:{$c, s$}] (t1f3) {$s_3$};
		\node[mNode,below=1cm of f1,label=left:{$\bar{c}, s$}] (f1t3){$s_5$};
		\node[mNode,right=1.6cm of f1t3, label=above:{$\bar{c}, \bar{s}$}] (f1f3){$s_6$};

		\node[mNode,below=1.4cm of t1f3, label=below:{$s, r$}] (t2) {$s_7$};
            \node[mNode,right=1.6cm of t2, label=right:{$s, \bar{r}$}] (t3){$s_8$};
		\node[mNode,below=1.4cm of f1f3, label=below:{$\,\,\,\,\,\,\,\,\,\bar{s}, \bar{r}$}] (f2) {$s_{10}$};
              \node[mNode,left=1.6cm of f2,label=right:{$\bar{s},r$}] (f3){$s_{9}$};
              
              	\node[mNode,label=left:{$w$}][below=2cm of t3] (s11) {$s_{11}$};
              \node[mNode,label=right:{$\bar{w}$}][below=2cm of f3] (s12){$s_{12}$};

		\draw [-{Latex[length=2mm]}] (dummy) -- (init) node[midway, above, sloped] {};
		\draw [-{Latex[length=2mm]}] (init) -- (t1) node[midway, above, sloped] {$0.5$};
		\draw [-{Latex[length=2mm]}] (init) -- (f1) node[midway, above, sloped] {$0.5$};
		
		\draw [-{Latex[length=2mm]}] (t1) -- (t1t3) node[midway,anchor=south west, yshift=5pt, right] {$0.9$};
		\draw [-{Latex[length=2mm]}] (t1) -- (t1f3) node[midway, above, sloped] {$0.1$};
		\draw [-{Latex[length=2mm]}] (f1) -- (f1t3) node[midway, anchor=south west, yshift=5pt, right] {$0.5$};
		\draw [-{Latex[length=2mm]}] (f1) -- (f1f3) node[midway, above, sloped] {$0.5$};	
		
		\draw [-{Latex[length=2mm]}] (t1t3) -- (f2) node[pos=0.85, above, sloped] {$0.2$}; //s4 ->
		\draw [-{Latex[length=2mm]}] (t1t3) -- (f3) node[pos=0.75, above, sloped] {$0.8$}; //s4 ->
		\draw [-{Latex[length=2mm]}] (t1f3) -- (t2) node[pos=0.25, left] {$0.8$}; //s3 ->
		\draw [-{Latex[length=2mm]}] (t1f3) -- (t3) node[pos=0.25, below, sloped] {$0.2$};	 //s3->
		\draw [-{Latex[length=2mm]}] (f1t3) -- (t3) node[pos=0.75, above, sloped] {$0.8$}; // s5->
		\draw [-{Latex[length=2mm]}] (f1t3) -- (t2) node[pos=0.85, above, sloped] {$0.2$}; //s5 ->
		\draw [-{Latex[length=2mm]}] (f1f3) -- (f3) node[pos=0.25, above, sloped] {$0.2$};
		\draw [-{Latex[length=2mm]}] (f1f3) -- (f2) node[pos=0.25, right] {$0.8$};

		\draw [-{Latex[length=2mm]}] (t2) -- (s11) node[pos=0.5, below, sloped] {$0.99$};
		\draw [-{Latex[length=2mm]}] (t2) -- (s12) node[pos=0.1, above, sloped] {$0.01$};
		\draw [-{Latex[length=2mm]}] (t3) -- (s11) node[pos=0.1, left] {$0.1$};
		\draw [-{Latex[length=2mm]}] (t3) -- (s12) node[pos=0.25, above, sloped] {$0.9$};	
		\draw [-{Latex[length=2mm]}] (f2) -- (s12) node[pos=0.5, below, sloped] {$1.0$};
		\draw [-{Latex[length=2mm]}] (f3) -- (s11) node[pos=0.15, above, sloped] {$0.9$};
		\draw [-{Latex[length=2mm]}] (f3) -- (s12) node[pos=0.25, right] {$0.1$};
		
		\path[every loop/.append style=-{Latex[length=2mm]}]
		(s11) edge [loop below] node {1} (s11)
		(s12) edge [loop below] node {1} (s12);
	\end{tikzpicture}
    }
	}
\end{minipage}

    \caption{(a) A pMC  and (b) one of its induced MCs.}
     \label{fig:mc-pmc}
 \end{figure}
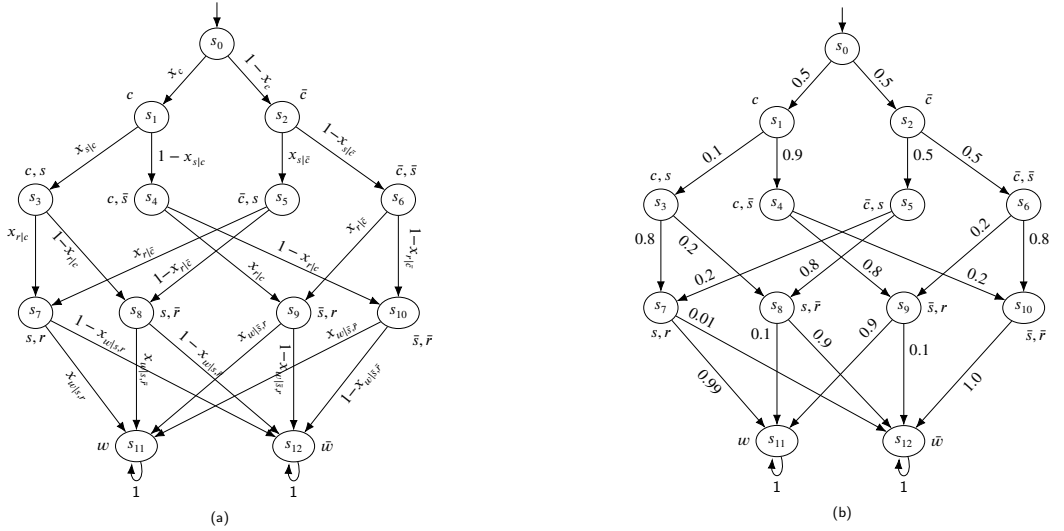
\begin{example} Figure \ref{fig:mc-pmc} (a) indicates a \emph{cycle-free} pMC with $13$ states.
 Figure \ref{fig:mc-pmc} (b) indicates its induced MC for the parameter instantiation $u$ with $u(x_c){=}u(x_{s| \bar c}){=}0.5$, $u(x_{s|c}) = u(x_{w|s, \bar r}){=}0.1$, $u(x_{r|c}){=}0.8$, $u(x_{r|\bar c}){=}0.2$, $u(x_{w|s, r}){=}0.99$, $u(x_{w|\bar s, r}){=}0.9$, and $u(x_{w|s, \bar r}){=}0$.
\end{example}

\par \noindent \paragraph{Computing reachability probabilities for parameter-free MCs. } The key procedure in verifying a MC against a temporal logic formula is computing reachability probabilities. Let $\mathcal{D} = (S, s_0, \mbox{Lits}, L, P)$ denote a MC.
Let $\mbox{\it Paths}(s)$ denote the set of all infinite paths in $\mathcal{D}$ starting from $s$, i.e., all infinite sequences of the form $\pi = s_1 s_2 s_3 \ldots$ with $s_1 = s$ and $P(s_i,s_{i+1}) > 0$. We denote the $i$'th state in path $\pi$ by $\pi(i)$.
A probability measure $\Pr_{\mathcal{D}}$ is defined on measurable sets of infinite paths using a standard cylinder construction; for details, see, e.g.,~\cite[Ch.~10]{BK08}.

\par \noindent Let $\mathcal{D}$ be a MC and $G \subseteq \mbox{Lits}$ be a set of goal atoms. Let $S_{G} \subseteq S$ be the set of goal states in $\mathcal{D}$: the states that satisfy $G$. For $\mathcal{D}$ and $s \in S$, the \emph{reachability probability} of $G$ from $s$ is the probability to eventually reach some state $g 
\in S_{G}$ and is denoted using standard notation from \emph{temporal logic} by $\Pr_{\mathcal{D},s}(\lozenge \, G )$ defined as follows.
\begin{equation}\label{eqn:probmeasure}
\Pr_{\mathcal{D},s}(\lozenge \, G) \ = \ \Pr_{\mathcal{D}} \{ \, \pi \in \mbox{\it Paths}(s) \mid \exists i. \, \pi(i) \in S_{G}  \, \}.
\end{equation}

Reachability probabilities can be obtained as the unique solution of a linear equation system~\citep{BK08} whose size is proportional to the number of states in $\mathcal{D}$.
\par \noindent \paragraph{The equation system for MC reachability probabilities. }
For the goal atoms $G$, let variable ${p_s}$ denote the probability to reach $S_{G}$ from state $s$. Variable $p_{s}$ is characterized as follows.
\begin{equation}
    p_s = \begin{cases}
        1, & \mbox{ if } s \in S_{G} \\
        0, & \mbox{ if } S_{G} \text{ is not reachable from } $s$, \text{ and}\\
		\underbrace{\sum\limits_{t \in S \setminus S_{G}} P(s,t) \cdot p_t}_{\text{reaching $G$ via state $t$}} + \underbrace{\sum\limits_{u \in S_{G}} P(s,u)}_{\text{reaching $G$ in one step}} & \mbox{ otherwise.}
    \end{cases}
\label{eq:equation-system}
\end{equation}
The above equation system yields the unique vector $\big( \Pr_{\mathcal{D},s}(\lozenge \, G) \big)_{s \in S}$. That is, we obtain the reachability of $G$ for all states $s$ by solving the equation system. We are often interested in the reachability probability from the initial state $s_0$ in $\mathcal{D}$ and we simply denote it as $\Pr_{\mathcal{D}}(\lozenge \, G)$.

\begin{example} 
\label{example-reach}
Consider the MC $\mathcal{D}$ in Figure \ref{fig:mc-pmc} (b). Let $G = \bar s$, that is we are interested in computing the probability of eventually reaching the literal $\bar s$ from the initial state $s_0$. We have $S_{G} = \{s_{4}, s_6\}$. For each state $s_{i}$, $p_{s_i}$ defines the probability of reaching $S_G$ from $s_i$ and is obtained by solving the following linear equation system:
\begin{align*}
& p_{s_0} = 0.5 \cdot p_{s_1} + 0.5 \cdot p_{s_2}, \\
& p_{s_1} = 0.1 \cdot p_{s_3} + 0.9 \cdot p_{s_4}, \\
& p_{s_2} = 0.5 \cdot p_{s_5} + 0.5 \cdot p_{s_6}.
\end{align*}
We also have $p_{s_{4}} = p_{s_6} = 1$ as $s_{4}, s_6 \in S_{G}$. Moreover, $p_{s_{3}} = p_{s_{5}} = \cdots = p_{s_{11}} = p_{s_{12}} = 0$ as $s_4$ and $s_6$ are not reachable from those states. Solving the equation system yields $p_{s_0} = \Pr_{\mathcal{D}}(\lozenge \, \bar s) = 0.7$.
\end{example}

\par \noindent \paragraph{Solution functions for pMCs. } 
For a parametric MC $\mathcal{M}$, $\Pr_{\mathcal{M}}(\lozenge \, G)$ is a function with $\Pr_{\mathcal{M}}(\lozenge \, G)[u] = \Pr_{\mathcal{D}}(\lozenge G)$ where $\mathcal{D} = \mathcal{M}[u]$, the so-called \emph{solution function}, see for details~\citep{DBLP:conf/ictac/Daws04}. 
\par \noindent Equation (\ref{eq:equation-system}) can be similarly defined for computing reachability objectives in pMCs. For pMCs, it yields a \emph{non-linear} equaion system its result amount to a rational function (polynomial in case of cycle-free MCs) rather than a concrete probability. See example below.

\begin{example}
    Consider the pMC $\mathcal{M}$ in Figure \ref{fig:mc-pmc} (a) and the same reachability formula $G = \bar s$ as in example \ref{example-reach}. The following system of non-linear equations is obtained.
\begin{align*}
& p_{s_0} = x_c \cdot p_{s_1} + (1-x_c) \cdot p_{s_2}, \\
& p_{s_1} = x_{s|c} \cdot p_{s_3} + (1 - x_{s|c}) \cdot p_{s_4}, \\
& p_{s_2} = x_{s|\bar c} \cdot p_{s_5} + (1 - x_{s|\bar c}) \cdot p_{s_6}.
\end{align*}
Similar to Example \ref{example-reach}, we have $p_{s_{4}} = p_{s_6} = 1$ and we have $p_{s_{3}} = p_{s_{5}} = \cdots = p_{s_{11}} = p_{s_{12}} = 0$. Solving the above equation system gives the following solution function: 
\begin{align*}
\Pr_{\mathcal{M}}(\lozenge \bar s) = p_{s_0} = x_{c} \cdot (1 - x_{s|c}) + (1-x_{c}) \cdot (1 - x_{s|\bar c}).
\end{align*}
\end{example}

\section{Mappings}
\label{sec:map}
We propose in this section (i) the mapping from cycle-free pMCs to ACs and (ii) the mapping from subclasses of ACs to pMCs.  
\subsection{Expressing Cycle-free pMCs as ACs}
In this section, we propose formal mappings from cycle-free pMCs to ACs and reduce computing reachability probabilities in such pMCs to a weighted model count in the obtained AC. This explains how the semantics of pMCs and ACs relate and how the notions in probabilistic model checking map to those in weighted model counting.


\par \noindent Let $\M = (S, s_0, X, \mbox{Lits}, L, P)$ be a pMC, where $S = \{s_1, \cdots, s_m \}$. Let $\mbox{\it succ}(s_i)$ denote the set of direct successor states for the state $s_i \in  S$, i.e., $\mbox{\it succ}(s_i) = \{s_j \in S \, \mid \,  P(s_i, s_j) > 0 \}$. 
We call $
\M$ a \emph{cycle-free} pMC iff it does not contain any path $\pi$ with $|\pi| > 1$ that $\pi(i) = \pi(j)$ for $i \neq j$, i.e., the pMC $\M$ is cycle-free iff it does not contain any backward transitions except for self-loops over single states.
{The mapping uses one sum node $n^+_i$ for each pMC state $s_i \in S$ and one multiplication node $n^\times_{i,j}$ for each transition $s_i {\rightarrow} s_j$. Thus, sum nodes represents states, while multiplication nodes represent transitions.}
Let $N_{S}^{+} = \{ n^{+}_{i} \, \mid \, s_i \in S \}$ denote the set of \emph{sum nodes}. 
Let $N_{S}^{\times} = \{n^{\times}_{i,j} \text{ for each } s_i, s_j \in S \text{ with } P(s_i, s_j) > 0 \}$ be the set of \emph{multiplication nodes}. 
Let $N_{\footnotesize \mbox{Lits}} = \{ n^{\alpha} 
 \text{ for each } \alpha \in \mbox{Lits}\}$ be the set of nodes corresponding to the literals, and $N_P$ 
 the set of nodes corresponding to distinct probability values in the transition probabilities. 


\begin{definition}[ACs of pMCs.]
\label{def:ac-of-mc}
Let $\M = (S, s_0, X, \mbox{Lits}, L, P)$ be a cycle-free pMC. The AC of $\M$ is the tuple $\mathcal{C}_{\M} = (N,n_0,\mbox{\it children},op,sym)$, where 
\begin{itemize}
    \item $N$ contains the inner nodes $N_S^{+} \cup N_S^{\times}$ and the leaf nodes $N_{\footnotesize \mbox{Lits}} \cup N_{P}$,
    \item the nodes in $N_S^{+}$ map to the operation $+$ by $op$ and the nodes in $N_{S}^{\times}$ map to $\times$,
    \item each \emph{sum node} $n^{+}_i$ corresponding to the state $s_i$ connects to the \emph{multiplication nodes} corresponding to $s_i$'s outgoing transitions, i.e., $ \mbox{\it children}(n^{+}_i) {=} \{n^{\times}_{i,j} \text{ for each } s_j {\in} \mbox{\it succ}(s_i) \}$,

    \item each \emph{multiplication node} $n^{\times}_{i,j}$ connects to the sum node $n^{+}_{j}$ that corresponds to the successive state $s_j$ (for $s_i \neq s_j)$, the probability node $n^p$ that is mapped to $p = P(s_i, s_j)$ by the mapping function $sym$, and the leaf nodes $n^{\alpha_1} \cdots n^{\alpha_k}$ that each map to a literal in $L(s_j)$. We thus have
   $\mbox{\it children}(n^{\times}_{i,j}) = \{ n^{+}_{j}\} \cup \{n^p\} \cup \bigcup\limits_{\alpha \in L(s_j)} n^{\alpha}$.
\end{itemize}

\end{definition}
\par \noindent The circuit $\mathcal{C}_{\M}$ includes an inner sum node for each state and an inner multiplication node for each pair of successive states in $\M$. Intuitively speaking, the sum node $n^+_i$ maps to the probability mass at state $s_i$ which is obtained by a sum over the probabilities of its branching paths. The probability of each branching path $\pi = s_i s_j \ldots$ is obtained by multiplying the transition probability $P(s_i, s_j)$ in the probability mass of $s_j$. This explains why $n_i^{+}$ connects to the multiplication nodes $n_{i,j}^{\times}$ for each successor $s_j$ and why $n_{i,j}^{\times}$ connects to $n_j^{+}$ and a leaf node $p$ with $sym(p) = P(s_i, s_j)$. The multiplication node $n_{i,j}^{\times}$ also connects to another set of leaf nodes $\{ n^{\alpha} \mid \alpha \in L(s_j)\}$ with $sym(n^{\alpha}) = \alpha$ that correspond to the literals at state $s_j$.
These leaf nodes allow computing reachability probabilities over the literals by a weighted model count on the obtained AC, see Proposition \ref{prop-mc-to-ac}.

\par \noindent Let $u: X \to \mathbb{R}$ be a \emph{well-formed instantiation} and $G = {l_1, \cdots, l_k} \subseteq \mbox{Lits}$ be a set of literals of goals to reach. We say the weight function $w: \mbox{Lits} \cup X \to \mathbb{Q}$ \emph{agrees} with $u$ iff $w(x) = u(x)$ for each $x \in X$. We say $w$ \emph{agrees} with $G$ iff
\vspace*{-0.1cm}
{{
\begin{align*}
    w(\alpha) = 1, w(\neg \alpha) = 0 \,\, \text{for each} \,\, \alpha \in G\quad \text{and} \quad w(\alpha) = w(\neg \alpha) = 1 \text{ for each } \,\, \alpha \not \in G.
\end{align*}
}
}
\begin{definition}[Path-compatible] A labeling function $L$ for the pMC $\M$ is \emph{path-compatible} iff for every pair of states $s, t$ in $\M$ with $t$ reachable from $s$, the labels of $s$ and  $t$ are compatible, that is, there is no literal $a$ such that $a 
\in L(s)$ and $\bar a \in L(t)$, or vica versa. The pMC $\M$ is path compatible iff its labeling is path-compatible.
\end{definition}

\begin{proposition}
\label{prop-mc-to-ac}
Let $\M = (S, s_0, X, \mbox{Lits}, L, P)$ be a cycle-free path-compatible pMC and $G \in \mbox{Lits}$ be a goal literal. Let $\mathcal{C}_{\M}$ be the AC of the MC $\M$ obtained from Def. \ref{def:ac-of-mc} and $w^{G}_{u}: X \cup \mbox{Lits}$ be a weight function that \emph{agrees} with $u$ and $G$. We then have
\begin{align*}
    \Pr_{\M[u]}(\lozenge \, G) = \mathcal{C}_{\M}[w^{G}_{u}].
\end{align*}
\end{proposition}

\begin{example}
    Consider the pMC $\mathcal{M}$ in Fig.~\ref{fig-mc-ac-updated} (b). Figure \ref{fig-mc-ac-updated} (a) indicates its AC $\mathcal{C}_{\M}$ obtained from Def.~\ref{def:ac-of-mc}. The AC starts with a root sum node that corresponds to the pMC's initial state. In general, each state in the pMC maps to a sum node in the obtained AC and each transition from $s$ to $s'$ maps to a multiplication node that is connected to (a) the labeling of state $s'$ and the transition probability $P(s, s')$ e.g., we have $P(s_0, s_1) = x_{c}$ in the pMC and $L(s_1) = c$, thus the multiplication node between the root node and the leftmost sum node (that is corresponding to $s_1$) is connected to the labeling $c$ and the weight variable $x_c$. The same holds for all other states and transitions. 
\end{example}

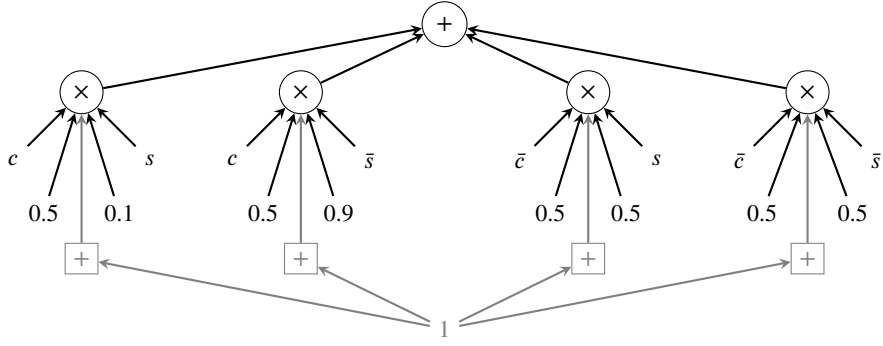
\begin{figure}
    \centering
    
\begin{tikzpicture} [node distance=0.9cm]
%

\node (n0) [plus] {$+$};

\node (n10) [times, below of=n0, left of=n0, xshift=-1cm] {$\times$};
\node (n11) [times, below of=n0, right of=n0, xshift=1cm] {$\times$};

\node (n12) [times, left of=n10, xshift=-2cm] {$\times$};
\node (n13) [times, right of=n11, xshift=2cm] {$\times$};

\node (n200) [leaf, below of=n12, left of=n12] {$c$};
\node (n201) [leaf, below of=n12, right of=n12] {$s$};
\node (n202) [leaf, below of=n12, left of=n12, xshift=0.4cm, yshift=-0.7cm] {$0.5$};
\node (n203) [leaf, below of=n12, right of=n12, xshift=-0.4cm, yshift=-0.7cm] {$0.1$};

\node (n210) [leaf, below of=n10, left of=n10] {$c$};
\node (n211) [leaf, below of=n10, right of=n10] {$\bar{s}$};
\node (n212) [leaf, below of=n10, left of=n10, xshift=0.4cm, yshift=-0.7cm] {$0.5$};
\node (n213) [leaf, below of=n10, right of=n10, xshift=-0.4cm, yshift=-0.7cm] {$0.9$};

\node (n220) [leaf, below of=n11, left of=n11] {$\bar{c}$};
\node (n221) [leaf, below of=n11, right of=n11] {$s$};
\node (n222) [leaf, below of=n11, left of=n11, xshift=0.4cm, yshift=-0.7cm] {$0.5$};
\node (n223) [leaf, below of=n11, right of=n11, xshift=-0.4cm, yshift=-0.7cm] {$0.5$};

\node (n230) [leaf, below of=n13, left of=n13] {$\bar{c}$};
\node (n231) [leaf, below of=n13, right of=n13] {$\bar{s}$};
\node (n232) [leaf, below of=n13, left of=n13, xshift=0.3cm, yshift=-0.7cm] {$0.5$};
\node (n233) [leaf, below of=n13, right of=n13, xshift=-0.3cm, yshift=-0.7cm] {$0.5$};

\node (n30) [draw=gray, text=gray, below= 1.7cm of n10] {$+$};
\node (n31) [draw=gray, text=gray, below = 1.7cm of n11] {$+$};
\node (n32) [draw=gray, text=gray, below = 1.7cm of n12] {$+$};
\node (n33) [draw=gray, text=gray,below = 1.7cm of n13] {$+$};

\node (n4) [leaf, text=gray,below = 3.5cm of n0] {$1$};
    
    \draw [dirarrow] (n10) -- (n0);
    \draw [dirarrow] (n11) -- (n0);
    \draw [dirarrow] (n12) -- (n0);
    \draw [dirarrow] (n13) -- (n0);

    \draw [dirarrow] (n200) -- (n12);
    \draw [dirarrow] (n201) -- (n12);
    \draw [dirarrow] (n202) -- (n12);
    \draw [dirarrow] (n203) -- (n12);

    \draw [dirarrow] (n210) -- (n10);
    \draw [dirarrow] (n211) -- (n10);
    \draw [dirarrow] (n212) -- (n10);
    \draw [dirarrow] (n213) -- (n10);

    \draw [dirarrow] (n220) -- (n11);
    \draw [dirarrow] (n221) -- (n11);
    \draw [dirarrow] (n222) -- (n11);
    \draw [dirarrow] (n223) -- (n11);

    \draw [dirarrow] (n230) -- (n13);
    \draw [dirarrow] (n231) -- (n13);
    \draw [dirarrow] (n232) -- (n13);
    \draw [dirarrow] (n233) -- (n13);

    \draw [gray,dirarrow] (n30) -- (n10);
    \draw [gray,dirarrow] (n31) -- (n11);
    \draw [gray,dirarrow] (n32) -- (n12);
    \draw [gray,dirarrow] (n33) -- (n13);

     \draw [gray,dirarrow] (n4) -- (n30);
    \draw [gray,dirarrow] (n4) -- (n31);
    \draw [gray,dirarrow] (n4) -- (n32);
    \draw [gray,dirarrow] (n4) -- (n33);
\end{tikzpicture}
  
    \caption{The AC in Fig.~\ref{fig:ac-example-small} equipped with artificial sum nodes and a $1$ leaf node.}
    \label{fig:ac-simple-equipped}
\end{figure}

\subsection{Expressing Arithmetic Circuits as pMCs } 
In this section, we express subclasses of arithmetic circuits (e.g., with alternating sum and product nodes and a sum node as the root) as pMCs, so that circuit evaluation reduces to computing a reachability solution function in the obtained pMC. 
\par \noindent Let $\mathcal{C} = (N, n_0, \mbox{\it children},op,sym)$ be an AC over the literals $Lits$ and the \emph{parameters} $X$ that is $sym: N_{L} {\to} (\mbox{Lits} \cup \mathbb{Q}[X])$.
\vspace{0.2cm}
\begin{remark}
    We consider the following assumptions for $\mathcal{C}$ in the rest of paper.
\begin{itemize}
\item We assume without loss of expressivity that $op(n_0) = +$; if the root node is instead a multiplication node $n^{\times}_0$, we equip the AC with a new sum node as the root and directly connect it to $n^{\times}_0$, see e.g., see Fig.~ \ref{fig:ac-independence} (a). 
\item We also assume without loss of expressivity that the type of each inner node in $\mathcal{C}$ alternates, that is, if $op(n) = +$, then each child node $n_{c} \in \mbox{\it children}(n)$ is either a leaf node or $op(n_{c}) = \times$. Similarly, if $op(n) = \times$, then each child node $n_{c} \in \mbox{\it children}(n)$ is either a leaf node or $op(n_{c}) = +$. 
\item At levels before the leaf nodes, the multiplication nodes are connected only to terminal nodes and not to any sum nodes. We connect each of such multiplication nodes to an \emph{artificial sum node\footnote{Note that the circuit evaluation is not modified by this modification. The idea is that the pMC states in Def.~\ref{def-pmc-of-ac} can be defined based on the AC's sum nodes.}} that itself connects to a leaf node labelled by $1$, see e.g., Fig.~\ref{fig:ac-simple-equipped}.
\item We restrict our mapping to ACs with \emph{probabilistic semantics} that are relevant to probabilistic inference. Let $n^+$ be sum node and $\{n^+_1, \cdots, n^+_k\}$ be the \emph{next} sum nodes of $n^+$, i.e., the sum nodes that are directly follows $n^+$ through some multiplication node: there is some $n^{\times}_i \in \mbox{\it children}(n^+)$ where $n^+_i \in \mbox{\it children}(n^{\times}_i)$. The AC $\mathcal{C}$ is \emph{probabilistic} iff (1) $< 0 \leq \mbox{\it weight}(n^{\times}_i) \leq 1$ and $\sum_{1 \leq i \leq k } \mbox{\it weight}(n^{\times}_i) \leq 1$.\footnote{For the parametric setting, this is ensured as we restrict to parametric probability distributions and well-formed instantiations.}
Note that we allow the sum of weights (that maps to sum of the outgoing transition probabilities from the corresponding state) to be below $1$ and not precisely $1$ as we redirect the remaining probability mass to the \emph{sink state}; see item (vi) in Def. ~\ref{def-pmc-of-ac}. This allows to cover ACs that are specifically for a given query or a given logical formula.  
\end{itemize}
\end{remark}
\par \noindent
We distinguish in our translation between the multiplication nodes that connect to a literal and those that do not: let $n^{\times} \in N$ and $op(n^{\times}) = \times$. We call $n^{\times}$ \emph{literal-equipped} iff there is some $n_l \in \mbox{\it children}(n^{\times})$ with $sym(n_l) \in \mbox{Lits}$. Such multiplication nodes carry over their literals as the labeling of their direct successors i.e., the states corresponding to their direct successive sum nodes; see item (iii) in Def. ~\ref{def-pmc-of-ac}.  We also distinguish between the multiplication nodes that are parent to only a single sum node and those that are parent to multiple sum nodes, see Fig.~ \ref{fig:ac-independence} (a). 
 The multiplication node $n^{\times}$ is \emph{simple} iff it is parent to a single sum node $n^{+} \in \mbox{\it children}(n^{\times})$. The multiplication node $n^{\times}$ is \emph{independence-derived} iff it is parent to more than a single sum node.
\par \noindent We define the \emph{weight} of a multiplication node as the polynomial (or the value) that is obtained by multiplying the values of its children that are non-literal leaf nodes. 


 \begin{figure}
     \centering
        \begin{tikzpicture}[
		node distance=0.3cm and 0.3cm,
		mNode/.style={draw,ellipse,align=center, minimum size=0.5cm},
		mLNode/.style={align=center, minimum size=0.5cm}
		]
		\node[](dummy){};
		\node[mNode,label=left:{}][below=0.5cm of dummy] (s0) {$s_0$}; 
		\node[mNode,label=left:{$c, s$}][below left=1.5cm and 4.5cm of s0] (s1) {$s_1$};
        \node[mNode,label=left:{$c, \bar s$}][below left=1.5cm and 1.2cm of s0] (s2) {$s_2$};
        \node[mNode,label=left:{$\bar c, s$}][below right=1.5cm and 4.55cm of s0] (s3) {$s_4$};
        \node[mNode,label=left:{$\bar c, \bar s$}][below right=1.5cm and 1.2cm of s0] (s4) {$s_3$};
         \node[mNode,label=left:{}][gray,text=gray,below=2.6cm of s0] (sink) {$\text{sink}$};

    	\draw [-{Latex[length=2mm]}] (dummy) -- (s0) node[pos=0.5, sloped, above] {};
		\draw [-{Latex[length=2mm]}] (s0) -- (s1) node[pos=0.5, sloped, above] {$0.5 \cdot 0.1$};
		\draw [-{Latex[length=2mm]}] (s0) -- (s2) node[pos=0.6,sloped, above] {$0.5 \cdot 0.9$};
        \draw [-{Latex[length=2mm]}] (s0) -- (s3) node[pos=0.5,sloped, above] {$0.5 \cdot 0.5$};
        \draw [-{Latex[length=2mm]}] (s0) -- (s4) node[pos=0.5,sloped, above] {$0.5 \cdot 0.5$};

         \draw [gray,-{Latex[length=2mm]}] (s1) -- (sink) node[gray,pos=0.5,sloped, above] {$1$};
        \draw [gray,-{Latex[length=2mm]}] (s2) -- (sink) node[gray,pos=0.5,sloped, above] {$1$};
         \draw [gray,-{Latex[length=2mm]}] (s3) -- (sink) node[gray,pos=0.5,sloped, above] {$1$};
        \draw [gray,-{Latex[length=2mm]}] (s4) -- (sink) node[gray,pos=0.5,sloped, above] {$1$};
        
	\path[every loop/.append style=-{Latex[length=2mm]}]
            (sink) edge [gray,loop below] node {1} (sink);

	\end{tikzpicture}
    \caption{The MC obtained from the AC in Fig.~\ref{fig:ac-simple-equipped} by Def.~\ref{def-pmc-of-ac}.}
    \label{fig:mc-of-simple-ac}
\end{figure}
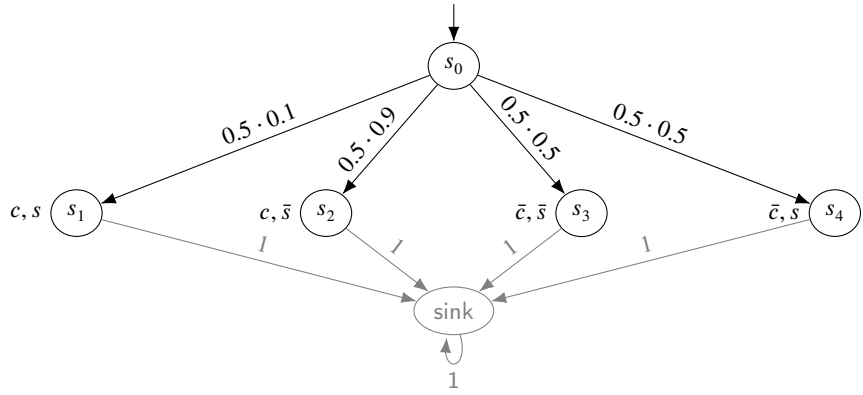

\begin{definition}[pMCs of ACs.] 
\label{def-pmc-of-ac}
The pMC of the AC $\mathcal{C}$ is the tuple $\mathcal{M}_{\mathcal{C}} = (S, s_0, X, \mbox{Lits}, L, P)$, where
\begin{enumerate}[label=(\roman*)]
    \item each sum node $n^{+}_i \in N$ with $op(n^{+}_i) = +$ yields a corresponding state $s_i \in S$.
    \item the root node $n_0$ yields the initial state $s_0$.
    \item For each state $s_i \in S \setminus \{s_0\}$ that corresponds to the sum node $n_i^{+} \in N$, the labeling of $s_i$ is defined as follows. Let $n_{\times}$ be \new{the parent} of  $n_i^{+}$ and $N_l$ be $n_i^{\times}$'s the literal nodes. We define $L(s_i) = \{op(n_l) \mid \in N_l \}$.
    \item Let $n^+_i, n^+_j \in N$ be two intermediate sum nodes that are connected through a \emph{simple} multiplication node $n^{\times}$ that is $n^{\times} \in \mbox{\it children}(n^+_i)$ and $n^+_j \in \mbox{\it children}(n^{\times})$. Let $s_i, s_j$ be the corresponding states to $n^+_i, n^+_j$. If $n^{\times}$ connects to the a child node $n_p$ with $op(n_p) = f \in \mathbb{Q}[X]$, then $P(s_i, s_j) = f$. Otherwise, $P(s_i, s_j) = 1$. For all other states $s, s'$ that their corresponding sum nodes are not directly connected through a multiplication node, $P(s, s') = 0$.
    \item Let $n^{\times}$ be an \emph{independence-derived} multiplication node, i.e., it is parent to several sum nodes $\{n^{+}_1, \cdots, n^{+}_k \}$ with $k > 1$. Let $n^{+}$ be $n^{\times}$'s parent. The following procedure is taken. For each of the sum node $n^{+}_i$, the sub-AC rooted at $n^{+}_i$ is translated to a sub-pMC $\mathcal{M}_i$ each rooted at the explicit state $s^i_{\times}$. The sub-pMCs are successively connected as follows. The state $s$ corresponding to $n^+$ ($n^{\times}$'s parent) connects with probability $1$ to $s^1_{\times}$ ($\mathcal{M}_i$'s initial state). This continues as follows. The states at the last level of $\mathcal{M}_{i-1}$ connect with probability $1$ to $s^i_{\times}$ until all sub-pMCs are connected.  Note that the order on $n^{+}_1$ to $n^{+}_k$ is arbitrary.
    \item The pMC $\mathcal{M}_{\mathcal{C}}$ is equipped with a sink state. Let $n^{+}_i$ be a sum node with the multiplication parents $n^{\times}_1, \cdots n^{\times}_k$, where $\sum_{1 \leq j \leq k} \mbox{\it weight}(n^{\times}_j) < 1$. The state $s_i$ corresponding to the sum node $n^+_i$ is then equipped with an additional transition to the \emph{sink state} with the transition probability $1 - \sum_{1 \leq j \leq k} \mbox{\it weight}(n^{\times}_j)$. The sink state absorbs the remaining probability mass for the states.    
\end{enumerate}
\end{definition}

\begin{example}
\label{example-mc-of-ac-simple}
    Figure \ref{fig:mc-of-simple-ac} indicates the MC \footnote{For simplicity, we showcase the mapping on an example with concrete probabilities rather than a parametric one.} obtained from the AC in Figure \ref{fig:ac-simple-equipped}. The root sum node maps to the MC's initial state. Each sum node (including the artificial sum nodes) maps to an MC state. The multiplication nodes and the probability values connected to them map to the transitions and the transition probabilities in MC. The literals at the multiplication node between the sum nodes $n^{+}_i$ and $n^{+}_j$ appear as the labeling of the state $s_j$ (corresponding to $n^{+}_j$). The leftmost state $s_1$ e.g., corresponds to the leftmost artificial sum node (shown in gray). It takes the labeling $c, \bar s$ that comes from the children of the left-most multiplication node. The transition probability $P(s_0, s_1) = 0.5 \times 0.1$ comes from the probabilities at this multiplication node. By the definition, the transition probabilities can also be arbitrary polynomials over the parameters $X$. In that case, the obtained model is a pMC with unknown parameters rather than only concrete probability values.
\end{example}

\begin{proposition} 
\label{prop-ac-to-mc}
Let $\mathcal{C}$ be an AC (with the assumptions set introduced in Remark 1), and $\mathcal{M}_{C}$ be its pMC obtained by Def. \ref{def-pmc-of-ac}. Let $G \in \mbox{Lits}$ be a goal literal and $w^{G}$ be a weight function that \emph{agrees with} $G$. Then
\begin{align*}
    \mathcal{C}[w^G] = \Pr_{\mathcal{M}_{\mathcal{C}}} (\lozenge \, G).
\end{align*}
\end{proposition}
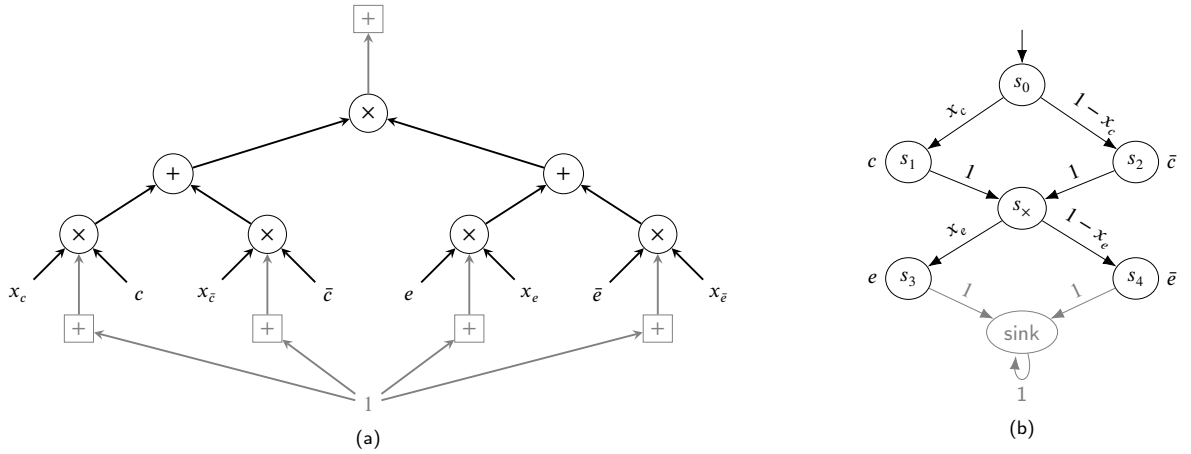
\begin{figure}
    \centering
     \begin{minipage}{0.63\linewidth}
    \resizebox{0.89\width }{0.89\height}{%
 \subfloat[]{
         \begin{tikzpicture} [node distance=0.9cm]
%

\node[gray] (n) [draw=gray,text=gray] {$+$};

\node (n0) [times,below of=n,yshift=-0.5cm] {$\times$};

\node (n10) [times, below of=n0, left of=n0, xshift=-2.cm] {$+$};
\node (n11) [times, below of=n0, right of=n0, xshift=2.cm] {$+$};

\node (n21) [times, below of=n10, left of=n10,xshift=-0.5cm] {$\times$};
\node (n22) [times, below of=n10,right of=n10,xshift=0.5cm] {$\times$};
\node (n23) [times, below of=n11, left of=n11,xshift=-0.5cm] {$\times$};
\node (n24) [times,below of=n11, right of=n11,xshift=0.5cm] {$\times$};

\node (n31) [leaf, below of=n21, right of=n21] {$c$};
\node (n311) [leaf, below of=n21, left of=n21] {$x_{c}$};
\node (n32) [leaf, below of=n22, right of=n22] {$\bar{c}$};
\node (n322) [leaf, below of=n22, left of=n22] {$x_{\bar{c}}$};

\node (n33) [leaf, below of=n23, left of=n23] {$e$};
\node (n333) [leaf, below of=n23, right of=n23] {$x_e$};

\node (n34) [leaf, below of=n24, left of=n24] {$\bar{e}$};
\node (n344) [leaf, below of=n24, right of=n24] {$x_{\bar{e}}$};

\node (n41) [draw=gray,text=gray,below of=n21, yshift=-0.5cm] {$+$};
 \node (n42) [draw=gray,text=gray,below of=n22, yshift=-0.5cm] {$+$};
 \node (n43) [draw=gray,text=gray,below of=n23, yshift=-0.5cm] {$+$};
 \node (n44) [draw=gray,text=gray,below of=n24, yshift=-0.5cm] {$+$};

 \node (n5) [leaf, text=gray,below = 3.8cm of n0] {$1$};

   \draw [dirarrow,gray] (n5) -- (n41);
    \draw [dirarrow,gray] (n5) -- (n42);
    \draw [dirarrow,gray] (n5) -- (n43);
   \draw [dirarrow,gray] (n5) -- (n44);

    \draw [dirarrow,gray] (n41) -- (n21);
    \draw [dirarrow,gray] (n42) -- (n22);
    \draw [dirarrow,gray] (n43) -- (n23);
    \draw [dirarrow,gray] (n44) -- (n24);

    \draw [dirarrow] (n31) -- (n21);
    \draw [dirarrow] (n311) -- (n21);

    \draw [dirarrow]  (n32) -- (n22);
    \draw [dirarrow]  (n322) -- (n22);

    \draw [dirarrow] (n33) -- (n23);
    \draw [dirarrow] (n333) -- (n23);

    \draw [dirarrow] (n34) -- (n24);
    \draw [dirarrow] (n344) -- (n24);

    \draw [dirarrow] (n21) -- (n10);
    \draw [dirarrow] (n22) -- (n10);
    \draw [dirarrow] (n23) -- (n11);
    \draw [dirarrow] (n24) -- (n11);

    \draw [dirarrow] (n10) -- (n0);
    \draw [dirarrow] (n11) -- (n0);

    \draw [gray,dirarrow] (n0) -- (n);
\end{tikzpicture}
}
}
        \end{minipage}
\hfill
 \begin{minipage}{0.31\linewidth}
    \resizebox{0.89\width }{0.89\height}{%
 \subfloat[]{
  \begin{tikzpicture}[
		node distance=0.3cm and 0.3cm,
		mNode/.style={draw,ellipse,align=center, minimum size=0.5cm},
		mLNode/.style={align=center, minimum size=0.5cm}
		]
		\node[](dummy){};
		\node[mNode,label=left:{}][below=0.5cm of dummy] (s0) {$s_0$}; 
        \node[mNode,label=left:{$c$}][below left=0.7cm and 1.2cm of s0] (s1) {$s_1$};
        \node[mNode,label=right:{$\bar{c}$}][below right=0.7cm and 1.2cm of s0] (s2) {$s_2$};

         \node[mNode,label=left:{}][below=1.2cm of s0] (smid) {$s_{\times}$};

        \node[mNode,label=left:{$e$}][below=1.1cm of s1] (s3) {$s_3$};
        \node[mNode,label=right:{$\bar{e}$}][below=1.1cm of s2] (s4) {$s_4$};
            
         \node[mNode,label=left:{}][gray,text=gray,below=1.2cm of smid] (sink) {$\text{sink}$};

    	\draw [-{Latex[length=2mm]}] (dummy) -- (s0) node[pos=0.5, sloped, above] {};
		\draw [-{Latex[length=2mm]}] (s0) -- (s1) node[pos=0.5, sloped, above] {$x_c$};
		\draw [-{Latex[length=2mm]}] (s0) -- (s2) node[pos=0.6,sloped, above] {$1-x_c$};
        \draw [-{Latex[length=2mm]}] (s1) -- (smid) node[pos=0.5,sloped, above] {$1$};
        \draw [-{Latex[length=2mm]}] (s2) -- (smid) node[pos=0.5,sloped, above] {$1$};
        \draw [-{Latex[length=2mm]}] (smid) -- (s3) node[pos=0.5,sloped, above] {$x_e$};
        \draw [-{Latex[length=2mm]}] (smid) -- (s4) node[pos=0.5,sloped, above] {$1-x_e$};
        
         \draw [gray,-{Latex[length=2mm]}] (s3) -- (sink) node[gray,pos=0.5,sloped, above] {$1$};
        \draw [gray,-{Latex[length=2mm]}] (s4) -- (sink) node[gray,pos=0.5,sloped, above] {$1$};
        
	\path[every loop/.append style=-{Latex[length=2mm]}]
            (sink) edge [gray,loop below] node {1} (sink);

	\end{tikzpicture}
    }
    }
\end{minipage}
    
   \caption{(a) The AC with a multiplication node that connects to several sum nodes, and (b) its corresponding pMC. }
    \label{fig:ac-independence}
\end{figure}
\begin{example} Assume that we are interested in the circuit evaluation $\mathcal{C}[w^G]$ for the AC $\mathcal{C}$ in Figure \ref{fig:ac-simple-equipped} and $G = \{\bar s\}$. The weight function $w^G$ that agrees with $G$ thus assigns $0$ to $s$ and assigns $1$ to other literals, i.e., $w^{G}(s)=0$, and $w^{G}(\bar s) = w^{G}(c) = w^{G}(\bar c) = 1$. We obtain the MC $\mathcal{M}_{\mathcal{C}}$ from $\mathcal{C}$; see Example \ref{example-mc-of-ac-simple} and Fig.\ref{fig:mc-of-simple-ac}.
We have by Proposition \ref{prop-ac-to-mc}
\begin{align*}
    \mathcal{C}[w^{\bar s}] = \Pr_{\mathcal{M}_{\mathcal{C}}} (\lozenge \, \bar s) = 0.5 \cdot 0.9 + 0.5 \cdot 0.5 = 0.7.
\end{align*}
\end{example}
\vspace*{-0.5cm}
\par \noindent We saw in Example \ref{example-mc-of-ac-simple} an AC (as well as its corresponding MC) in which each multiplication node connects only to a single sum node. We show in Example \ref{example-independence-ac} an AC in which a multiplication nodes connect to several sum nodes and illustrate how the (p)MC is constructed for such case. Such ACs mostly originate from examples in which the random variables are \emph{independent} and thus the probabilities of the independent variables can be simply multiplied to obtain the joint probability distributions.

\begin{figure}
    \centering
     \makebox[\textwidth][c]{
    \begin{minipage}{0.3\linewidth}
    \resizebox{0.95\textwidth}{!}{
        \subfloat[]{{     \centering
        \begin{tikzpicture}[
		node distance=0.3cm and 0.3cm,
		mNode/.style={draw,ellipse,align=center, minimum size=0.5cm},
		mLNode/.style={align=center, minimum size=0.5cm}
		]
		\node[](dummy){};
		\node[mNode,label=left:{}][below=0.5cm of dummy] (s0) {$s_0$}; 
        \node[mNode,label=left:{$c$}][below left=0.7cm and 1.2cm of s0] (s1) {$s_1$};
        \node[mNode,label=right:{$\bar c$}][below right=0.7cm and 1.2cm of s0] (s2) {$s_2$};


        \node[mNode,label=left:{$e$}][below=1.3cm of s1] (s3) {$s_3$};
        \node[mNode,label=right:{$\bar{e}$}][below=1.3cm of s2] (s4) {$s_4$};
            
         \node[mNode,label=left:{}][gray,text=gray,below=3.3cm of s0] (sink) {$\text{sink}$};

    	\draw [-{Latex[length=2mm]}] (dummy) -- (s0) node[pos=0.5, sloped, above] {};
		\draw [-{Latex[length=2mm]}] (s0) -- (s1) node[pos=0.5, sloped, above] {$0.5$};
		\draw [-{Latex[length=2mm]}] (s0) -- (s2) node[pos=0.6,sloped, above] {$0.5$};
         \draw [-{Latex[length=2mm]}] (s1) -- (s3) node[pos=0.5,sloped, above] {$0.0001$};
        \draw [-{Latex[length=2mm]}] (s1) -- (s4) node[pos=0.75,sloped, above] {$0.9999$};
        \draw [-{Latex[length=2mm]}] (s2) -- (s3) node[pos=0.75,sloped, above] {$0.0001$};
        \draw [-{Latex[length=2mm]}] (s2) -- (s4) node[pos=0.5,sloped, above] {$0.9999$};  
      
         \draw [gray,-{Latex[length=2mm]}] (s3) -- (sink) node[gray,pos=0.5,sloped, above] {$1$};
        \draw [gray,-{Latex[length=2mm]}] (s4) -- (sink) node[gray,pos=0.5,sloped, above] {$1$};
        
	\path[every loop/.append style=-{Latex[length=2mm]}]
            (sink) edge [gray,loop below] node {1} (sink);

	\end{tikzpicture}
 }}%
    }
    \end{minipage}%
    \begin{minipage}{0.3\linewidth}
    \resizebox{0.95\textwidth}{!}{
        \subfloat[]{{     \centering
        \begin{tikzpicture}[
		node distance=0.3cm and 0.3cm,
		mNode/.style={draw,ellipse,align=center, minimum size=0.5cm},
		mLNode/.style={align=center, minimum size=0.5cm}
		]
		\node[](dummy0){};
		\node[][below=.5cm of dummy0] (dummy){};
		\node[mNode,label=left:{}][below=0.5cm of dummy] (s0) {$s_0$}; 
        \node[mNode,label=left:{$c$}][below left=0.85cm and 1.2cm of s0] (s1) {$s_1$};
        \node[mNode,label=right:{$\neg c$}][below right=0.85cm and 1.2cm of s0] (s2) {$s_2$};

         \node[mNode,label=left:{}][below=1.7cm of s0] (smid) {$[s_{3}]$};

         \node[][below=1.5cm of smid] (dummy2) {};

    	\draw [-{Latex[length=2mm]}] (dummy) -- (s0) node[pos=0.5, sloped, above] {};
		\draw [-{Latex[length=2mm]}] (s0) -- (s1) node[pos=0.5, sloped, above] {$0.5$};
		\draw [-{Latex[length=2mm]}] (s0) -- (s2) node[pos=0.6,sloped, above] {$0.5$};
        \draw [-{Latex[length=2mm]}] (s1) -- (smid) node[pos=0.5,sloped, above] {$1$};
        \draw [-{Latex[length=2mm]}] (s2) -- (smid) node[pos=0.5,sloped, above] {$1$};
        
        
	\path[every loop/.append style=-{Latex[length=2mm]}]
            (smid) edge [gray,loop below] node {1} (smid);
	\end{tikzpicture}}}%
    }
    \end{minipage}
    \begin{minipage}{0.3\linewidth}
    \resizebox{0.95\textwidth}{!}{
        \subfloat[]{{     \centering
        \begin{tikzpicture}[
		node distance=0.3cm and 0.3cm,
		mNode/.style={draw,ellipse,align=center, minimum size=0.5cm},
		mLNode/.style={align=center, minimum size=0.5cm}
		]
		\node[](dummy0){};
		\node[][below=.5cm of dummy0] (dummy){};
		\node[mNode,label=left:{}][below=0.5cm of dummy] (s0) {$[s_0]$}; 


       \node[mNode,label=left:{$e$}][below left=0.7cm and 1.2cm of s0](s3) {$s_3$};
       \node[mNode,label=right:{$\neg e$}][below right=0.7cm and 1.2cm of s0] (s4) {$s_4$};
            
         \node[mNode,label=left:{}][gray,text=gray,below=1.7cm of s0] (sink) {$\text{sink}$};

        \node[][below=1.7cm of smid] (dummy2) {};

    	\draw [-{Latex[length=2mm]}] (dummy) -- (s0) node[pos=0.5, sloped, above] {};
        \draw [-{Latex[length=2mm]}] (s0) -- (s3) node[pos=0.5,sloped, above] {$0.0001$};
        \draw [-{Latex[length=2mm]}] (s0) -- (s4) node[pos=0.5,sloped, above] {$0.9999$};
        
         \draw [gray,-{Latex[length=2mm]}] (s3) -- (sink) node[gray,pos=0.5,sloped, above] {$1$};
        \draw [gray,-{Latex[length=2mm]}] (s4) -- (sink) node[gray,pos=0.5,sloped, above] {$1$};
        
	\path[every loop/.append style=-{Latex[length=2mm]}]
            (sink) edge [gray,loop below] node {1} (sink);
\vspace{0.4cm}
	\end{tikzpicture}}}%
    }
    \end{minipage}
    }
   \caption{(a) {The MC related to the AC in Fig.~\ref{fig:ac-independence} (b) by Def.~\ref{def-pmc-of-ac}, (b) its bisimulation quotient w.r.t. the literals $\{c, \neg c\}$, and (c) its bisimulation quotient w.r.t the literals $\{e, \neg e\}$.}}
   \label{fig:mc-independence-triple}
\end{figure}
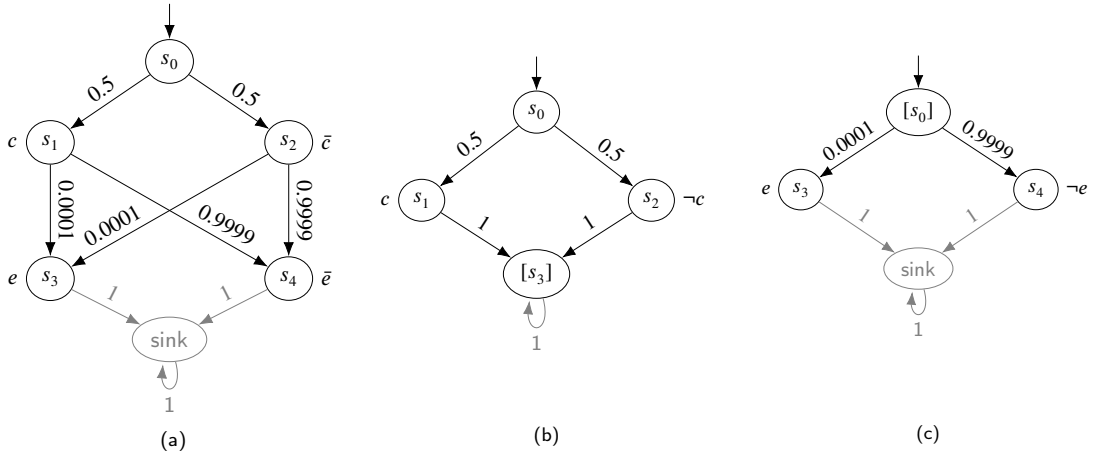

\begin{example} 
\label{example-independence-ac}
 Let us consider the literals $c$, $\bar c$ that as before represent \emph{cloudy} and \emph{not cloudy} as well as the literals $e$ and $\bar e$ that represent \emph{earthquake} and \emph{no earthquake}. Figure \ref{fig:ac-independence} (a) indicates an AC over these literals, where the random variables for \emph{cloudy} and \emph{earthquake} are \emph{independent}: the probabilities for $C$ and $E$ can be simply multiplied to compute $\Pr(C \wedge E)$. We see that in this AC, a multiplication node connects to several sum nodes, one of which encodes the probability distribution for the random variable $C$ and the other encodes the probability distribution for the random variable $E$. The (p)MC's transition probability function maps each state to a (parametric) probability distribution over its successors; therefore, several independent distributions cannot directly connect to a single state (the initial state in this example): the subgraphs at the multiplication node are thus handled successively ---and recursively if the subgraph itself includes a multiplication node with several sum nodes. This means an arbitrary order over the sum nodes is taken; the first subgraph is translated to pMC states and transitions up to the leaf nodes. The last states for this subgraph are then connected to the state corresponding to the second sum node (i.e., the root of the second subgraph). The procedure continues until all the multiple sum nodes are taken care of. Figure \ref{fig:ac-independence} (b) indicates the corresponding pMC for the AC in Fig. ~\ref{fig:ac-independence} (a). The initial state $s_0$ corresponds to the artificial root (sum) node in the AC. The multiplication node is connected to two sum nodes that need to be handled successively. First, the left subgraph of the AC is handled that yields the states $s_1$ and $s_2$. Then the right subgraph is handled that yields the states $s_3$ and $s_4$. Note that the intermediate stated $s_{\times}$ is considered for illustration purposes to explicitly indicate the multiplication semantics. It can be removed and the states $s_3$ and $s_4$ can be directly connected to $s_1$ and $s_2$ without affecting the reachability probabilities.  
\end{example}

\paragraph{Probabilistic Bisimulation and Probabilistic Independence. }A key optimization technique in \emph{probabilistic model checking} is
\emph{bisimulation minimization}: the model is minimized based on \emph{probabilistic bisimulation} \citep{BK08}, which is the probabilistic counterpart to classical bisimulaton. The idea to partition the states with \emph{identical behaviour} and ultimately obtain the coarsest sub-model that is relevant to the properties of interest. 

We illustrate how bisimulation minimization relates to independence in the following example.

\begin{example}
\label{example-bisim-ind} Recall Example \ref{example-independence-ac}, where we showcased a very simple example with only two independent variables. This can correspond to a Bayesian network (BN) with two disconnected nodes for the random variables $E$ (earthquake) and $C$ (cloudy). 
Figure \ref{fig:mc-independence-triple} (a) indicates a Markov chain related to this example, where the multiplication node is removed (without changing the underlying probability distributions) and the parameters are instantiated for better illustration. Figure \ref{fig:mc-independence-triple} (b) shows the bisimulation quotient w.r.t. the literals of the random variables \emph{Cloudy} and Figure \ref{fig:mc-independence-triple} (c) for the random variable \emph{Earthquake}. Intuitively speaking, when one of the independent random variables does not appear in the query (and is irrelevant) or when the conditioning makes the random variables conditional-independent, bisimulation minimization gives the \emph{sub-model} with only relevant states. 
\end{example}

\section{Discussion}
\label{sec:discussion}
\par \noindent In this section, we complete our formal mappings with a comparative discussion between weighted model counting and probabilistic model checking. Table \ref{tab:wmc-pmc-inference} summarizes how notions from the latter correspond to those from the former. 

\par \noindent Inference probabilities, e.g., for Bayesian networks are computed as \emph{reachability probabilities} of MCs in probabilistic model checking \citep{DBLP:conf/qest/SalmaniK20,DBLP:conf/ecsqaru/SalmaniK21,DBLP:journals/jair/SalmaniK23}, whereas weighted model counting computes the queried probabilities by traversing and evaluating the AC \citep{DBLP:journals/ai/ChaviraD08}. Markov chains are state-based models, where each state forms a probability distribution over its successors, whereas ACs are computational models that only encode how probability distributions are encoded as polynomials using the sum, multiplication, and leaf nodes. Generally, it is not necessarily required that the intermediate nodes in the ACs yield probability distributions. However, in our mapping from ACs to pMCs, we assume that in such cases the weights are normalized. Valuation of random variables map to literals (over atomic propositions \footnote{Note that the original papers on Bayesian inference using probabilistic model checking, the states are explicitly defined over the joint variable valuations without being labeled by atomic propositions or literals. We take literals to ensure consistency between probabilistic model checking definitions and those in weighted model counting.}) both in the pMC and the AC. In weighted model counting, this notion was originally referred to as \emph{indicator variables}: the indicator variable $\lambda_{v_i=d}=0$ iff $v_i = d$ violates the evidence. This maps to assigning the state variables $p_{s_i}$ to $0$ when the goal (the evidence) is not reachable from the states; see Equation (\ref{eq:equation-system}) and Example \ref{example-reach} on page \pageref{example-reach}. 
In the MC, the probability entries, e.g., from the conditional probability tables (CPTs) of the Bayesian networks are stored as transition probabilities, whereas in the AC, the CPT entries are stored as weights directly connected to the multiplication nodes. The repeated weights are possible to be shared in the AC. While the complexity of computation is primarily determined by the number of edges in the circuit, sharing the weights still allows the models with many repeated probabilities to be represented more succinctly. While in Markov chains, the probability distributions over independent variables cannot simply be multiplied as in arithmetic circuits (the so-called factorization), preprocessing based on graph analysis and bisimulation minimization enable efficient computation of the reachability probabilities in such cases; see, e.g., Example \ref{example-bisim-ind}. It has been experimentally shown in previous studies \citep{DBLP:conf/qest/SalmaniK20} that the initial graph analysis in probabilistic model checking often pays off as it automatically trims the irrelevant states.

\begin{table}[!htb]
    \centering
    
  \begin{tabular}
   {|c|c|c|}\hline
 \makebox[10em]{Probabilistic Model Checking}&\makebox[10em]{Weighted Model Counting} & \makebox[10em]{Probabilistic inference} \\\hline\hline
 Markov chain & Arithmetic Circuit & E.g., Bayesian network \\\hline
 computing reachability probabilities & AC evaluation & computing inference probabilities \\\hline
 literals (over atomic propositions) & literals (over atomic propositions)  & (single) variable valuation\\\hline
  transition probabilities & AC weights & CPT entries (for BNs)\\\hline
 probability of the paths & terms & probability of joint valuations \\\hline
probabilistic bisimulation & factorization  & probabilistic independence \\\hline
\end{tabular}

        \caption{Weighted model counting vs. probabilistic model checking: notions and entities related to probabilistic inference.}
    \label{tab:wmc-pmc-inference}
\end{table}

\par \noindent Table \ref{tab:wmc-pmc} overviews some key capabilities of weighted model counting and probabilistic model checking at a glance. 
Probabilistic model checking inherently supports non-determinism on the modelling level as Markov decision processes (as well as extensions with partial observability). WMC by itself cannot represent non-deterministic choices, since each assignment gets a single numeric weight. One can encode some non-deterministic behaviour into WMC by adding extra choice variables and then analyzing bounds e.g., computing min/max WMC over the instantiations. Support for computing expected rewards exists in both frameworks. Using WMC one can compute expected rewards by changing the operators to algebraic ones. Rather similarly, in probabilistic model checking the states can be equipped with rewards and the expected values can be computed for reachability probabilities or for arbitrary PCTL formulas: the rewards are collected upon leaving each state. 
\par \noindent Existing experimental results \citep{DBLP:journals/jair/SalmaniK23}that contrast WMC and PMC show that for non-cyclic Bayesian networks and simple inference queries e.g., with simple unary hypotheses WMC can be significantly faster for some \emph{very large} models. This can be explained through the direct computational model with the possibility of sharing repeated entries in ACs and better variable ordering heuristics. Circuit-based inference methods for WMC, however, have limitations in models and queries they support. A key capability of probabilistic model checking is its inherent support for cyclic models that enables inference in more complex models e.g., for dynamic Bayesian networks \citep{DBLP:conf/atva/PalaniappanT12} or probabilistic programs with loops. Moreover, the native support for linear temporal logic as the query language in probabilistic model checking allows flexibility  ---without the need for separately compiled models for a specific query. This is also useful for extensions of Bayesian networks such as dynamic Bayesian networks to pose queries on the trajectories of the events and on arbitrary points in time.
\begin{table}[!b]
    \centering
    \begin{tabular}
   {|c|c|c|}\hline
\makebox[12em]{} &\makebox[12em]{Probabilistic Model Checking}&\makebox[12em]{Weighted Model Counting}\\\hline\hline
direct support for non-determinism &  \checkmark & \xmark\\\hline
support for computing expected rewards &  \checkmark & \checkmark \\\hline
support for cyclic models & \checkmark & \xmark \\\hline
flexible query language by linear temporal logic & \checkmark & \xmark \\\hline
efficient inference for large models & \xmark & \checkmark \\\hline
\end{tabular}
        \caption{Weighted model counting vs. probabilistic model checking: capabilities. 
        }
    \label{tab:wmc-pmc}
\end{table}

\section{Conclusion and Future Directions}

We proposed a mapping from cycle-free parametric Markov chains to Arithmetic circuits that reduces computing reachability probabilities on the MC to a weighted model count on the obtained AC. We showed how probabilistic model checking and weighted model counting contrast/relate for probabilistic inference. We also briefly examined probabilistic bisimulation on the Markov chains of Bayesian networks, mirroring probabilistic independence — similar to how factorization captures independence in weighted model counting.

Our results contribute to multiple directions for research. First, for sparse, acyclic MCs, arithmetic circuits may offer a more compact and operationally efficient representation than traditional transition matrices. This could potentially have implications for the scalability of model checking in large, sparse systems. Second, we propose extending our mapping to the computation trees of cyclic MCs for the purpose of computing finite-horizon reachability probabilities similar to \citep{DBLP:conf/cav/HoltzenJVMSB20}. 
\par \noindent In the context of parameter synthesis for parametric Markov chains (pMCs), our approach suggests new directions for efficiently computing solution functions. Notably, by encoding both the pMC and its polynomial parametrized transitions as arithmetic circuits, one might leverage symbolic computation to accelerate synthesis. The experimental results in \citep{DBLP:conf/atva/GainerHS18} have previously shown the effectiveness of such encoding to mitigate memory consumption for storing very large rational functions. Moreover, the transfer of parameter synthesis techniques for pMCs to arbitrary probabilistic circuits becomes feasible using our AC2pMC mapping. Finally, for Bayesian networks and Dynamic Bayesian networks \citep{DBLP:conf/atva/PalaniappanT12}, our mapping facilitates the translation of heuristics—such as variable ordering strategies—between probabilistic model checking and weighted model count settings. Our bidirectional study is expected to improve inference and parameter synthesis efficiency across both domains, particularly in models that combine probabilistic reasoning with logical structure.

\vspace*{0.7cm}
\par \noindent \textbf{Acknowledgements. } The authors would like to thank Joost-Pieter Katoen for his contributions and valuable input on the earlier papers that led to this work. We are also grateful to Luc De Raedt for his valuable input and the technical discussions that inspired us to work on this report. Finally, we thank the anonymous reviewers of \cite{DBLP:journals/jair/SalmaniK23} for emphasizing the importance of a comparative formal study between weighted model counting and probabilistic model checking.


\bibliographystyle{cas-model2-names}

\bibliography{sample}

\end{document}